\documentclass[aps,pre,superscriptaddress,floatfix,tightenlines,showpacs,twocolumn,10pt,notitlepage]{revtex4-2}
\usepackage{amsmath}
\usepackage{amssymb}
\usepackage{amsthm}
\usepackage{bm}
\usepackage{bbm}
\usepackage[shortlabels]{enumitem}
\usepackage{tabularx}
\usepackage{graphicx}
\usepackage{tikz-cd}
\usepackage{xcolor}
\usepackage{hyperref}
\usepackage[normalem]{ulem}
\usepackage[textsize=scriptsize,backgroundcolor=white]{todonotes}
\usepackage{url}
\usepackage[dvipsnames,svgnames,table]{xcolor}
\usepackage{framed}
\usepackage{color} 
\usepackage{comment}
\usepackage{mathtools}

\hypersetup{
	colorlinks=true,
	linkcolor=blue,
	citecolor=blue,			%Crimson
	urlcolor=blue,			%Crimson
	filecolor=blue,
	anchorcolor = green,
}

\newcommand{\gf}[1]{\textbf{\textcolor{RoyalBlue}{Gary: #1}}}

\newcommand{\aleks}[1]{\textbf{\textcolor{violet}{Aleks: #1}}}
\newcommand{\aleksadd}[1]{\textcolor{violet}{#1}}

\theoremstyle{definition}

\newtheorem{algorithm}{Algorithm}

\newcommand{\Prob}{{\mathrm{Prob}}}

\begin{document}
\title{Transient almost-invariant sets reveal convective heat transfer patterns in plane-layer Rayleigh-B\'{e}nard convection}

\author{Aleksandar Badza}
\affiliation{School of Mathematics and Statistics, UNSW Sydney, Sydney NSW 2052, Australia}

\author{Gary Froyland}
\affiliation{School of Mathematics and Statistics, UNSW Sydney, Sydney NSW 2052, Australia}

\author{Roshan J. Samuel}
\affiliation{Institute of Thermodynamics and Fluid Mechanics, Technische Universit\"at Ilmenau, P.O.\ Box 100565, D-98684 Ilmenau, Germany}

\author{Jörg Schumacher}
\affiliation{Institute of Thermodynamics and Fluid Mechanics, Technische Universit\"at Ilmenau, P.O.\ Box 100565, D-98684 Ilmenau, Germany}

\date{\today}

\begin{abstract}
Horizontally extended plane-layer convection flows are characterized by characteristic patterns of turbulent heat transfer due to the convective fluid motion consisting of a nearly-regular ridge network where hot fluid rises and cold fluid sinks. 
Here, we analyse this transport behavior by the so-called inflated generator framework, which identifies quasi-stationary families of almost-invariant sets, derived from leading inflated generator eigenvectors. 
We demonstrate the effectiveness of this data-driven analysis framework in three-dimensional turbulent flow, by extracting  transient characteristic heat transfer patterns as families of almost-invariant sets subject to a transient evolution, which contribute least to the convective heat transfer.
\end{abstract}

\maketitle

%\tableofcontents

\section{Introduction}
Turbulent convection phenomena are ubiquituous starting with heat and momentum transfer processes in stellar interiors \cite{Garaud2021} via vortex formation in planetary atmospheres to magnetic field generation by dynamo action in planetary cores \cite{Landeau2022} and heat exchanger and cooling devices in technology \cite{Smolentsev2010}. In the simplest setting buoyancy forces drive fluid motion, caused by the temperature dependence of the fluid density, in a planar fluid layer of height $H$ that is uniformly heated from below ($z=0$) and uniformly cooled from above ($z=H$) \cite{Koschmieder1993,Verma2018}. This is the fundamental Rayleigh-B\'{e}nard convection (RBC) case \cite{Rayleigh1916,Chilla2012} in which the mass density is a linear function of the temperature deviation from a reference equilibrium state, the working fluid is incompressible, and the material parameters, such as thermal conductivity and kinematic viscosity, are constant across the fluid volume. For imposed temperature differences between top and bottom, which significantly exceed a critical threshold value, a (turbulent) fluid motion is initialized. Heat is transferred then by diffusion and dominantly by advection. This thermal driving is quantified by the dimensionless Rayleigh number $Ra$ \cite{Koschmieder1993,Verma2018}.    

One central question in turbulent convection is the exact amount of heat that is transferred on average from the bottom to the top as a function of the imposed outer temperature difference (i.e., on $Ra$) and the material properties of the working fluid in the plane layer \cite{Kadanoff2001,Ahlers2009,Chilla2012}. Depending on the horizontal extension (in $x$ and $y$ directions) of the fluid layer, this turbulent transport proceeds frequently in the form of regular and gradually evolving horizontal fluid flow patterns at scales that exceed the thickness $H$ of the layer. These patterns set a large-scale order, which exists despite the convection flow is fully turbulent on smaller scales; they are termed {\it turbulent superstructure} of convection \cite{Pandey2018,Stevens2018,Alam2025}. Recently, it has been shown that these structures are rooted in a hierarchical (i.e.\ successively coarser) network of thermal plumes with increasing distance from the wall.
These plumes are linearly unstable fragments of the thermal boundary layer that detach from the walls and rise into the bulk. This implies that turbulent superstructures are formed in a bottom-up process, starting with the smallest building blocks --the thermal plumes-- from the walls \cite{Shevkar2025}. In addition, it is found that the characteristic ``mesh width'' of this plume network follows a self-similar scaling with Rayleigh number, which scales with the Rayleigh number as $\sim Ra^{-1/3}$.      

The large-scale patterns of plane-layer convection depend on the thermal boundary conditions \cite{Vieweg2021}. For prescribed temperature at the top and bottom, the Dirichlet boundary condition case (which will be considered here), we observe a collection of horizontally curved and elongated circulation roll structures with horizontal cross-roll modulations to form circulation cells which fill the layer vertically from bottom to top. They also exist as a single entity in closed cylindrical convection cells with cell diameter of the order of cell height and are termed here large-scale circulation (LSC) \cite{Kadanoff2001,Ahlers2009,Shi2012}. 

In terms of the material transport in the turbulent fluid, i.e., from the perspective of the Lagrangian frame of reference, these convection cells can be considered as coherent sets \cite{Froyland2010} -- regions that largely trap Lagrangian tracer particles for a transient period of time in their cores and thus essentially reduce the mixing and heat transfer from bottom to top in the convection case. This concept has been discussed in several other material transport problems in the atmosphere and ocean \cite{Froyland2010,GHRSS12, Allshouse2015,FPG15}. The complementary perspective is to detect and study the manifolds between the nearly coherent sets, which can serve as barriers to material transport \cite{FPG14,Haller2015,DKF25}. Both frameworks stem from dynamical systems theory. Coherent sets were investigated in RBC in different ways, e.g., by various implementations of the dynamic Laplacian or dynamic clustering \cite{Schneide2018,Schneide2019,Klunker20,Vieweg2021a,Schneide2022}, their long-term behavior was described by a Markov model \cite{Maity2022}, the arising advective and diffusive transport barriers were analysed \cite{Aksamit2023}, and so-called collective variables were calculated to model transitions between large-scale states \cite{Maity2023}. 

Despite turbulent convection flows being time-dependent, turbulent superstructures \textit{slowly form and only gradually change} before fading away. While they exist, they are therefore {\it almost-invariant sets} \cite{DJ99} i.e., they stay approximately fixed in space and only slowly leak fluid. Due to this quasi-stationary-in-space property of the turbulent superstructures, we analyse the convective heat transfer (CHT) in a turbulent plane-layer RBC flow using almost-invariant sets rather than coherent sets. In contrast to standard almost-invariant set analysis, which can only detect almost-invariant sets that possess the almost-invariance property throughout the flow duration, here we study the flow over long duration in which we expect the birth and death of many superstructures. We seek \emph{quasi-stationary families of almost-invariant sets}, namely time-dependent fluid parcels that may appear, slowly evolve in a quasi-stationary-in-space manner, and ultimately disappear. Each such family represents a convection cell, as discussed above. We go beyond prior work by  tracking the onset and decay of the convection cells and filtering out possibly spurious coherent sets that move rapidly in the fluid.

We employ a recent data-driven framework \cite{BF24}, built around the (time-varying) infinitesimal generators $G_t$, which generate the Perron-Frobenius or transfer operators, the operators which describe the evolution of the distribution of a (probability) measure in the phase space of a dynamical system for a positive time. We ``inflate'' (time-extend) the individual generators ${G_t}$ from the spatial into the spatio-temporal domain of the time-dependent RBC flow in the plane layer to create an inflated generator ${\cal G}$. Through this special time-extension we are able to capture the almost-invariant character of the transport patterns, as well as their appearance, time evolution, and disappearance. 

An eigenvector analysis of ${\cal G}$, which builds on the Sparse Eigenbasis Approximation (SEBA) \cite{FRS19} technique is subsequently used to identify quasi-stationary families of almost-invariant sets. Our analysis can extract the gradually evolving patterns in the bulk of the horizontally extended plane-layer turbulent convection flow; they can be successfully compared with the time-windowed Eulerian fields of the convective heat transfer, which correlates velocity and temperature and thus forms the important transport pattern in the RBC flow \cite{Fonda2019}. We note that velocity data are used in the present framework to reproduce correlations between velocity and temperature. We thus provide a proof of concept to apply the framework from dynamical systems theory to reveal the spatio-temporal organization of convective heat transfer in horizontally extended plane-layer configurations, which are termed mesoscale convection flows \cite{Alam2025}, from reduced information. The analysis will also allow one to analyse the correlation of the SEBA fields with $u_z$.

The outline of the manuscript is as follows. Section II introduces the convection flow with its parameters. Here, we also detail the convective heat transfer, the coupled field of velocity and temperature, which characterizes the turbulent heat transfer due to fluid motion. Section III summarizes the analysis methods for almost invariant sets. Afterwards, in Sec. IV, we present the results of our data analysis before the final conclusions are drawn in Sec. V.  

\section{Rayleigh-B\'{e}nard convection model}
\subsection{Boussinesq equations and parameters}
The data for the analysis are generated by direct numerical simulations (DNS). To this end, we solve the three-dimensional Navier-Stokes equations in the Boussinesq approximation numerically. This couples velocity ${\bm u}({\bm x},t)=(u_x,u_y,u_z)$ and temperature $T({\bm x},t)$ fields with ${\bm x}=(x,y,z)$~\citep{Chilla2012,Verma2018}. The non-dimensional form of the governing equations is given by 
\begin{eqnarray}
\frac{\partial {\bm u}}{\partial t} + ({\bm u} \cdot {\bm \nabla}) {\bm u} & = & -{\bm \nabla}p + T \hat{{\bm z}} + \sqrt{\frac{Pr}{Ra}} \, \nabla^2 {\bm u}, \label{eq:u} \\
\frac{\partial T}{\partial t} + ({\bm u} \cdot {\bm \nabla}) T & = & \frac{1}{\sqrt{PrRa}} \, \nabla^2 T, \label{eq:T} \\
{\bm \nabla} \cdot {\bm u} & = & 0. \label{eq:m}
\end{eqnarray}
The equations are solved by the spectral element method (SEM) using the GPU accelerated SEM solver, nekRS~\citep{Fischer2022}. The computational domain is a cuboidal cell of $\Gamma =L/H= 8$ with periodic boundary conditions along the horizontal directions, where $L$ is the horizontal extension and $H$ is the height. The horizontal coordinates are $x,y \in [-4,4]$. The top and bottom walls are found at $z=0$ and $1$. The simulation domain is the volume $V=L^2 H$.

We set no-slip condition on the velocity field and prescribe  uniform temperatures $T_{\rm top}= 0$ and $T_{\rm bot} = 1$ respectively. 
The two dimensionless parameters that characterize the convection flows (in addition to the aspect ratio $\Gamma$) are the Rayleigh number $Ra$ and the Prandtl number $Pr$, which are given by
\begin{equation}
\label{eq:Ra}
Ra=\frac{g\alpha\Delta TH^3}{\nu\kappa} \quad \mbox{and} \quad Pr=\frac{\nu}{\kappa}\,,
\end{equation}
where $g$ is the acceleration due to gravity, $\Delta T=T_{\rm bot}-T_{\rm top}>0$ is the temperature difference between the hot and cold plates, $\alpha$ is the thermal expansion coefficient, $\kappa$ is the thermal diffusivity, and $\nu$ is the kinematic viscosity. The present study uses one data set at $Ra=10^5$ and $Pr=0.7$. The characteristic length, velocity and temperature scales used to obtain the non-dimensionalized eqns. \eqref{eq:u}--\eqref{eq:m} are $H$, $U_f=\sqrt{g\alpha\Delta T H}$, and $\Delta T$ respectively, where $U_f$ is the free-fall velocity of the thermal plumes.
%---------------------------------------
\begin{figure*}
\centering
\includegraphics[width=0.95\textwidth]{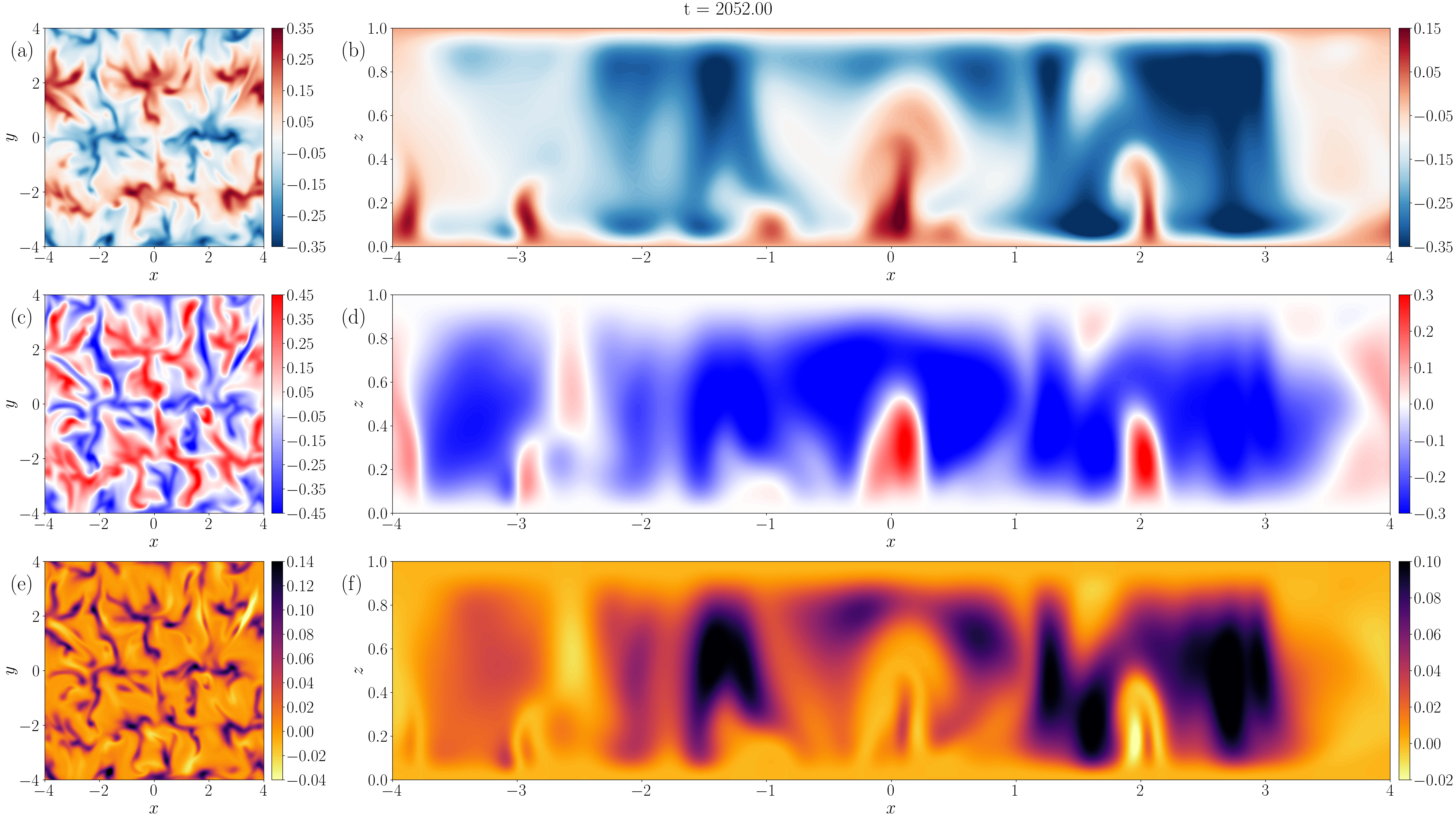}
      \caption{Snapshots of the temperature fluctuation $\theta$, in panels (a,b), the vertical velocity $u_z$, in panels (c,d), and convective heat transfer field in panels (e,f) for the time instant at $t=2052 \,T_f$. The left column displays horizontal cuts at $z=1/2$, the right column vertical cuts at $y=0$.}
\label{fig:rbc_cht}
\end{figure*}
%---------------------------------------

We follow the same workflow as in Samuel et al. \cite{Samuel2024}. Consequently, we ensure that the meshes are fine enough to resolve the boundary layers. The simulation s with the diffusive equilbrium state of RBC for which the fluid is at rest and the temperature drops linearly from the bottom to the top. This equilibrium state is infinitesimally perturbed and develops into a turbulent convection flow for a transient period. The turbulent state is statistically stationary, i.e., statistical moments and correlations are time-independent. Time averages, indicated by $\langle\cdot\rangle_t$, are taken in the following always as an arithmetic average over a sequence of statistically independent snapshot realizations of the fluid flow. Time averages will be combined with averages over the horizontal cross section plane $A=L^2$ or the simulation domain $V=AH$ and denoted by $\langle\cdot\rangle_{A,t}$ and $\langle\cdot\rangle_{V,t}$, respectively. Error bars are then standard deviation from this statistical analysis.

We determined Nusselt number, $Nu$, and Reynolds number, $Re$, which are the global measures of heat and momentum transport, respectively. To this end, we compute $Nu$ from the non-dimensionalized temperature gradient at the bottom wall, whereas $Re$ is calculated from the volume-averaged root mean square (rms) velocity, see also eq. \eqref{eq:Ra},
\begin{equation}
Nu=-\frac{\partial \langle T\rangle_{A,t}}{\partial z}\bigg|_{z=0},\quad
Re=U_{\rm rms} \sqrt{\frac{Ra}{Pr}}\,.
\label{NuRe}
\end{equation}
The root mean square velocity is computed for the volume $V$, $U_{\rm rms}=\langle |{\bm u}|^2\rangle_{V,t}^{1/2}$. 

To solve the system of equations~\eqref{eq:u} to~\eqref{eq:m} using SEM, we discretize the domain into $N_e = 300 \times 300 \times 64$ elements. Each element is further discretized into Gauss-Lobatto-Legendre (GLL) nodes from fifth-order ($p=5$) Legendre polynomials along each direction, yielding a total of $N_e \times p^3 =$ 720 million collocation points. 

Each snapshot generated from DNS contains velocity, pressure and temperature data at all collocation points. We obtained 150 such snapshots within steady-state, separated by unit $T_f$, where $T_f = H/U_f$ is the free-fall time used to non-dimensionalize time , covering a total of 150 $T_f$ (resulting in 3.6 TB of data). The free-fall time $T_f$ is a characterisitic time that estimates how long cold dense fluid falls across a distance $H$ under the action of buoyancy forces. To ensure a statistically steady state, data was sampled after an initial period of 2000 $T_f$. Note that although the DNS is performed at sufficiently high spectral resolution -- necessary to accurately calculate the temporal evolution of the velocity and temperature fields -- the datasets are subsequently downsampled for the detection of the quasi-stationary almost-invariant sets.
%---------------------------------------
\begin{figure*}
\centering
\includegraphics[width=0.7\textwidth]{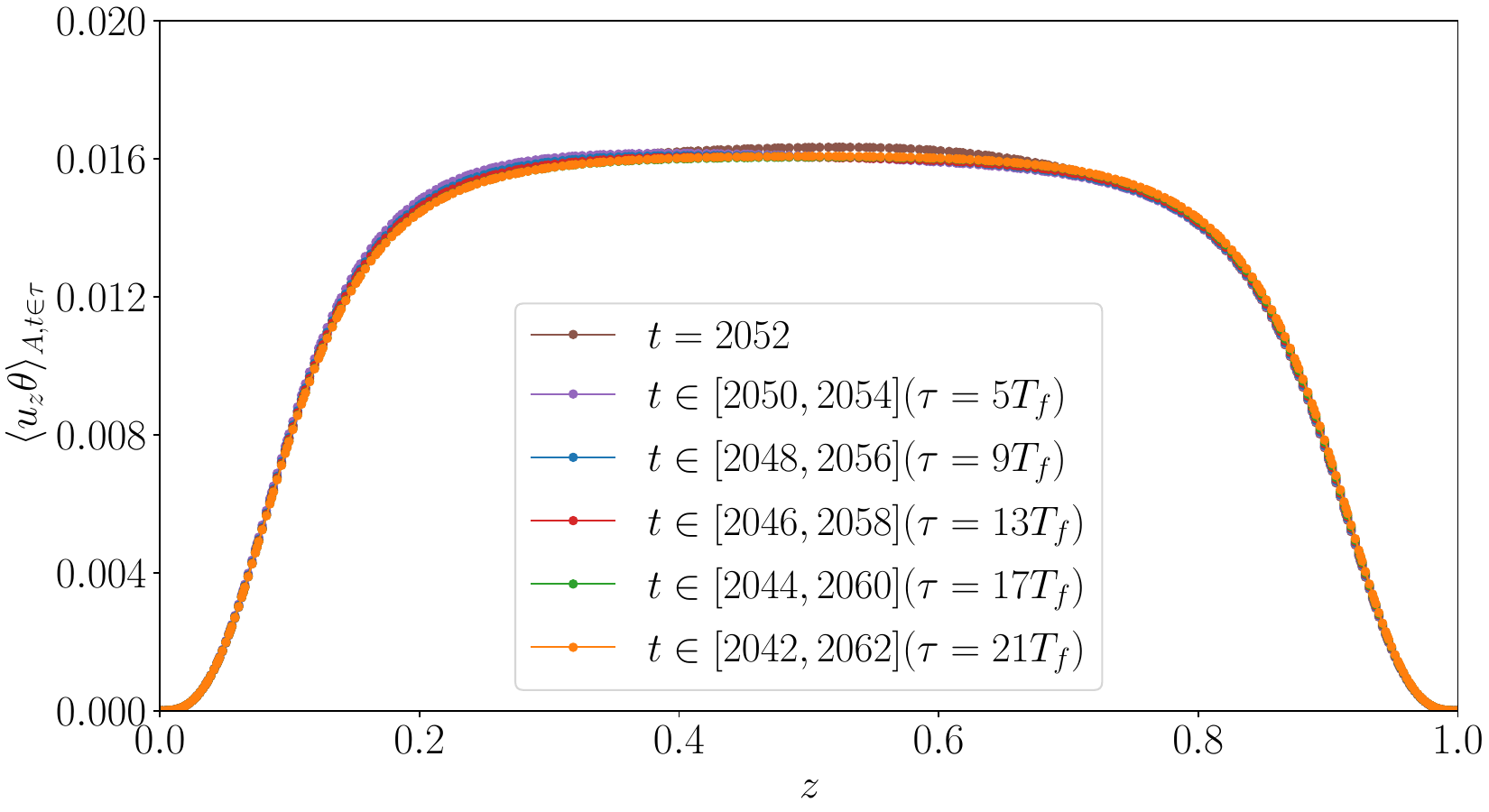}
      \caption{Mean values of convective heat transfer (CHT) across the horizontal plane at different $z$-levels plotted against these $z$ levels, at $t=2052 T_f$, and averaged over time windows of length 5, 9, 13, 17 and 21 $T_f$, centred at time $t=2052 T_f$.  Similar results are obtained at other times.}
\label{fig:rbc_cht_planemeans}
\end{figure*}
%---------------------------------------

Averaging over a sequence of statistically independent snapshots in the statistically stationary regime, we obtain the Nusselt number from our DNS to be $Nu = 4.264 \pm 0.058$, and the Reynolds number is $Re = 92.66 \pm 1.03$. These values are consistent with those obtained by Samuel et al. \cite{Samuel2024}. The slightly reduced value of $Re$ is also consistent with the observations made for larger aspect ratios as shown for $Ra=10^9$ in the same work.

\subsection{Convective heat transfer field}
While the global heat transport is quantified in non-dimensional units using the Nusselt number, a local measure can be obtained the convective heat transfer (CHT) field, which is given by
\begin{equation}
H({\bm x},t) = u_z({\bm x},t)\theta({\bm x},t)\,,
\label{CHT}
\end{equation}
where $u_z({\bm x},t)$ is the component of the velocity field along the vertical $z$-direction, and 
\begin{equation}
\label{cht}
\theta({\bm x},t) = T({\bm x},t) - \langle T \rangle_{A,t}(z)
\end{equation}
the temperature fluctuation, obtained after subtracting the mean temperature profile
$\langle T \rangle_{A,t}(z)$. Here, $\langle \cdot\rangle_{A,t}$ is a combined average over the horizontal cross section $A=L^2$ and time $t$. Using \eqref{CHT}, it is possible to calculate the Nusselt number in an alternative way to \eqref{NuRe} which is given by 
\begin{equation}
Nu = 1 + \sqrt{Ra Pr} \langle H ({\bm x},t) \rangle_{V,t}\,.
\end{equation}
Computing $Nu$ using this formula, we get $Nu = 4.264 \pm 0.086$, which is in excellent agreement with the former value $Nu$ computed from near-wall temperature gradient \eqref{NuRe}. This underlines, that the numerical simulations are well-resolved, both in the bulk of the layer and the near-wall regions. 

We obtain a sequence of 102 three-dimensional data snapshots of the velocity and temperature fields of RBC flow for times $2001 T_f < t < 2103 T_f$ with a time interval of 1 $T_f$ between successive snapshots. The field data are subsequently  interpolated spectrally to a coarser three-dimensional grid with uniform mesh size $\ell$ in all three space directions, see also Sec. III C. The coarse grid will be used for the data analysis. Note that the original fields are much more finely resolved with a vertically non-uniform spectral element grid, in particular close to the bottom and top walls. 

Figure \ref{fig:rbc_cht} shows two-dimensional slice cuts through the temperature field $T({\bm x},t_0)$ in panels (a,b), the vertical velocity field $u_z({\bm x},t_0)$ in panels (c,d), and the CHT field $H({\bm x},t_0)$ in (e,f) at $t_0=2052 T_f$, which is the snapshot in the middle of the sequence. In the left column of the figure a horizontal cross section at the midplane at $z=1/2$ is shown, in the right column we show a corresponding vertical cut through the volume $V$ at $y=0$. While the horizontal cuts display the spatial organization of the convection patterns by means of three fields viewed from the top, the vertical cut highlights the thermal plume structures of the convection flow across the layer which carry the heat from bottom to top.  

It is seen that CHT is locally maximal where hot fluid rises and cold fluid sinks to the bottom; in both cases locally $H>0$. Note, that in both cases, heat is carried from the bottom to the top. For the vertical cut at $y=0$, this is dominantly done by cold fluid falling down as seen by relating panel (a) with panel (f) of Fig. \ref{fig:rbc_cht}. These fields fluctuate strongly with respect to space and time. We thus smoothen out the snapshot data by a sliding time average over a time window $[t_k - m, t_k + m]$ of length $\tau=(2m+1) \ T_f$ for the upcoming analysis. We will take $m = 2$ and thus a time window of length $5 \ T_f$. This filtering with respect to time, which will be denoted by $\langle\cdot\rangle_{\tau}$ highlights the large-scale and gradually evolving convection cells better, which we aim to analyse with the present framework. 

While the short-time fluctuations of the fields are smoothened out by the sliding average, mean transport properties of the flow remain unaffected by this procedure. To demonstrate this, we choose a single snapshot at a time $t_0$ and calculate the mean vertical CHT profile $\langle H(t_0)\rangle_A (z)$. This is compared in Fig. \ref{fig:rbc_cht_planemeans} with the mean profiles that are obtained from an average which combines the time window-average centered about $t_0$ and plane average of CHT fields, $\langle H\rangle_{A, \tau}(z)$.  
We see that the vertical mean profiles overlap for differently long time windows $\tau$, which are indicated in the legend. All CHT profiles are $H=0$ at the walls, which is caused by the no-slip boundary condition at the walls, $z=0,1$, which states that $u_x=u_y=u_z=0$. The averaged CHT profiles become maximal in the center or bulk of the convection layer as seen in Fig. \ref{fig:rbc_cht_planemeans}.

\section{Analysis method for quasi-stationary families of almost-invariant sets}

As discussed in the introduction, we aim to characterise the three-dimensional fluid motion using the concept of  \textit{quasi-stationary families of almost-invariant sets} \cite{BF24}. For our turbulent RBC flow, we expect many such families to be present simultaneously, although each individual family may not exist throughout the full time duration we consider from $2001T_f$ to $2103T_f$. The \emph{almost-invariant} aspect of these families means that at any given time instant, almost all of the fluid remains in the family over short time durations.
The \emph{quasi-stationary} aspect relates to the families being approximately fixed in the spatial domain through time, and also allows for families to emerge and disappear. These families are the manifestation the slow dynamics of the large-scale turbulent superstructure patterns, the spatial complement of the families will be responsible for the major fraction of the CHT across the plane layer.  

We use the approach of \cite{BF24} to construct multiple quasi-stationary families of almost-invariant sets.
This approach rests on a spectral analysis of generators for stochastic drift-diffusion equations, where the drift is given by $\bm{u}(t,\bm x)$.
\begin{equation}
    \label{SDE_Static}
    \mathrm{d} \bm x_s = \bm u(t, \bm x) \ \mathrm{d}s + \epsilon \ \mathrm{d} \bm w_s
\end{equation}
where $\{\bm w_s\}$ is Brownian motion and $\epsilon > 0$ is small.

% , namely emerging and dissipating static (or almost static) subsets within the flow domain over a finite period of time. The second method uses convective heat transfer to quantify the change of temperature throughout the system which aids in the movement of fluid throughout the system.
% \gf{Currently the CHT method is completely missing from section 3.}

\subsection{Almost-invariant sets and generators of steady flows}

%We begin by briefly defining the concept of quasi-stationary families of almost-invariant sets and the inflated generator method used to identify these sets. We then describe the computational implementation of this method for a non-autonomous flow defined in three-dimensional space (such as the RBC flow), and then talk about isolating specific quasi-stationary families of almost-invariant sets through sparse eigenbasis approximation (SEBA).

%Here, we briefly summarise the concept of quasi-stationary families of almost-invariant sets and the inflated generator method used to identify these families of sets \gf{I could not find the brief summary...where is it?}. For a more comprehensive description of the underlying mathematical theory behind these concepts, we refer the reader to relevant mathematical literature, e.g. \cite{BF24,FJK13}.

%Suppose we have a flow system defined within a domain $M \subset \mathbb{R}^3$ with velocity $\bm u$ \gf{Is this $u$ fixed or time-varying?}. At a particular fixed moment in time $t$, we add a small stochastic perturbation to the velocity $\bm u$ \gf{we add Brownian motion continuously in time, not just a fixed moment} which gives us a stochastic version of our velocity system

This section provides background on identifying almost-invariant sets in \emph{steady} flows, i.e.\ where $\bm{u}(t,x)=\bm{u}(x)$;  in the next section, we extend these ideas to time-dependent velocity fields.
For a steady vector field $\bm{u}$, we consider trajectories $\{\bm{x}_t\}$ of the SDE \eqref{SDE_Static}.
We say that $S \subset M$ is \emph{almost-invariant} if
\begin{equation}
    \label{rho_Static}
\rho^s(A):=\frac{\Prob(\bm x_0\in A, \bm x_{s}\in A)}{\Prob(\bm x_0\in A)}\approx 1
\end{equation}
for short durations $s$, where $\Prob$ is the probability of the event occurring if $\bm{x}_0$ is initialised uniformly over $M$.

Following \cite{FJK13}, one may construct a first-order-in-$s$ formula for $\rho^s(S)$: 
\begin{equation}
    \label{rho_prop_3.2}
    \rho^{s} (A) = 1 + \frac{G_{A} \mathbf{1}_{A}}{V(A)} \cdot s + o(s),
\end{equation}
where $V$ denotes volume in $M$ and $\mathbf{1}_{A}$ is the indicator function on $A$.
The operator $G_A$ is defined as 
\begin{equation}
    \label{gen_functional}
    G_{A} f := \lim\limits_{s \rightarrow 0} \int\limits_{A} \frac{P^s f - f}{s} \ \mathrm{d}V,
\end{equation}
where $P^s$ is the Perron--Frobenius operator, describing evolution of a density $f$ by the Fokker--Planck equation 
\begin{equation}
    \label{fokker-planck}
    \frac{\partial f(t,\bm x)}{\partial t} = - \nabla \cdot (\bm u (\bm x) f(t,\bm x)) +  \frac{\epsilon^2}{2} \Delta f(t,\bm x)
\end{equation}
for a duration $s$.
From \eqref{gen_functional} it is clear that $G_A\mathbf{1}_A\le 0$ and returning to \eqref{rho_prop_3.2}, we see that almost-invariant sets $A$ should have $G_{A} \mathbf{1}_{A} \lessapprox 0$.
To obtain multiple almost-invariant sets $A$, following \cite{FJK13}, one computes several leading real eigenfunctions of the generator $G$, where
 \begin{equation}
     \label{gen_pf_limit}
     Gf := \lim\limits_{s \rightarrow 0} \frac{P_t^s f - f}{s}.
 \end{equation}
That is, we solve the linear eigenproblem 
$Gf=\lambda f$ 
%,\qquad \nabla f(\bm x)\cdot \bm n(\bm x)=0\mbox{ for }\bm x\in\partial M,
with periodic boundary conditions applied at the vertical ``sides'' of $M$ and Neumann boundary conditions applied at the floor and ceiling of $M$.
%(though a similar problem can be defined using Dirichlet or Periodic boundary conditions), and use the ``leading" eigenfunctions $f_k$ (i.e. those corresponding to eigenvalues $\lambda_k < 0$ sufficiently close to 0) to identify almost-invariant sets present within $M$ at time $t$. 

\subsection{Quasi-stationary families of almost-invariant sets and the inflated generator}
In this and subsequent sections we assume that $\bm{u}$ is time-varying.
We define the \textit{spacetime} domain $\mathbb{M} = [0,\tau] \times M$, which should be thought of as a continuum of copies of $M$ across times from 0 to $\tau$.
We now define spacetime versions of $\bm{u}$, the SDE in \eqref{SDE_Static}, the invariance ratio $\rho^s$, and the generator $G$.

Following \cite{BF24} we define a time-expanded
domain $\mathbb{M} = [0, \tau]\times M$ and construct a stochastic process as so that:
\begin{enumerate}
    \item 
 in the time coordinate, introduce diffusion with intensity $a > 0$,
\item in the original phase space coordinates, on $M$ at time $t$, we have the usual SDE on $M$, namely diffusion with intensity $\epsilon$ and drift given by $\bm{u}(t, ·)$.
\end{enumerate}
More precisely, we define a spacetime vector field $\bm{U}$ by
\begin{equation*}
    \bm U(t,\bm x)=\begin{pmatrix}
     0\\
     \bm u(t,\bm x)  
\end{pmatrix}
\end{equation*} 
and a spacetime SDE as 
\begin{equation}
    \label{SDE_Dynamic}
    \mathrm{d}\bm X_s = \bm U(\bm X_s) \ \mathrm{d}s + \bm \Sigma \ \mathrm{d}\bm W_s,
\end{equation}
where $\bm X = (t,\bm x)\in \mathbb{M}$ and 
\begin{equation*}
    \bm \Sigma=\begin{pmatrix}
    a & 0 \\ 0 & \epsilon \bm I_3
\end{pmatrix}.
\end{equation*}
The scalar $\epsilon > 0$ continues to represent diffusion in space, and $a > 0$ is the diffusion coefficient along the temporal coordinate. 
Note that \eqref{SDE_Dynamic} is an \textit{autonomous} SDE on $\mathbb{M}$.
%a new parameter representing diffusion in time.
%Using the matrix $\bm \Sigma$, we are extending purely spatial diffusion over $M$ as defined in \eqref{SDE_Static}, to spatiotemporal diffusion (governed by the two parameters $\epsilon$ and $a$) across our spacetime domain $\mathbb{M}$. 

Similarly to the previous subsection, we will say that the spacetime set $\mathbb{A}\subset\mathbb{M}$ is almost-invariant if
\eqref{rho_Static}, 
\begin{equation}
    \label{rho_Dynamic}
\rho^s(\mathbb{A}):=\frac{\Prob(\bm X_0\in \mathbb{A}, \bm X_{s}\in \mathbb{A})}{\Prob(\bm X_s\in \mathbb{A})} \approx 1,
\end{equation}
where $\bm{X_0}$ is distributed uniformly on $\mathbb{M}$.
Let's write the spacetime set $\mathbb{A}=\bigcup_{t\in [0,\tau]} (\{t\}\times A_t)$, where each $A_t\subset M$, $t\in[0,\tau]$ and interpret what almost-invariance in spacetime means.
Stochastic trajectory escape from a spacetime set $\mathbb{A}$ can occur in the space coordinates or the time coordinate, and if $\rho^s(\mathbb{A})\approx 1$, both types of escape should be low.
\begin{itemize}
    \item 
Low escape in space coordinates means that on most time-slices $\{t\}\times A_t$, one has $\rho^s(A_t)\approx 1$, and therefore at most time instances $A_t$ is \textit{almost-invariant in the usual phase space sense}.
\item Low escape in the time coordinate means that the boundary of $\mathbb{A}$ is mostly aligned with the time axis, which means that there is \textit{small variation of the boundary of the family $A_t$ with $t$};  in other words, \textit{the family $\{A_t\}_{t\in[0,\tau]}$ is quasi-stationary}.
\end{itemize}
Note that the sets $A_t$ may be empty for during some subintervals of $[0,\tau]$.
This flexibility allows for quasi-stationary families of almost-invariant sets to \textit{appear and disappear in time}.

% Thus, almost-invariant sets 
% which satisfies an updated version of the property 
% \begin{enumerate}
%     \item $\mathbb{S}$ is comprised of individual sets $S_t$ which are almost-invariant in space at their respective moments in time $t \in [0,\tau]$,
%     \item The entrance and escape rates for particle volume between sets $S_t$ defined on adjacent time slices are as small as possible between these time slices under the effect of temporal diffusion, allowing the full family of these sets contained in $\mathbb{S}$ to remain quasi-stationary over time, and
%     \item The sets $S_t$ do not have to be non-empty for the entire time interval, and may be non-empty only for $[t_1,t_2] \subset [0,\tau]$.
% \end{enumerate}
Following \cite{BF24}, one may construct a first-order-in-$s$ formula for $\rho^s(\mathbb{A})$: 
\begin{equation}
    \label{rho_prop_3.2_infgen}
    \rho^{s} (\mathbb{A}) = 1 + \frac{G_{\mathbb{A}} \mathbf{1}_{\mathbb{A}}}{\mathbb{V}(\mathbb{A})} \cdot s + o(s),
\end{equation}
where $\mathbb{V}$ denotes volume in $\mathbb{M}$. 
An operator $\mathcal{G}_{\mathbb{A}}$ is defined as 
\begin{equation}
    \label{gen_functional-spacetime}
    \mathcal{G}_{\mathbb{A}} F := \lim\limits_{s \rightarrow 0} \int\limits_{\mathbb{A}} \frac{\mathcal{P}^s F- F}{s} \ \mathrm{d}\mathbb{V},
\end{equation}
where $\mathcal{P}^s$ is the Perron--Frobenius operator, describing evolution of a density $F$ by the Fokker--Planck equation 
\begin{equation}
    \label{fokker-planck-spacetime}
    \frac{\partial F(\bm X)}{\partial s} = - \nabla \cdot (\bm U (\bm X) F(\bm X)) +  \frac{\Sigma^2}{2} \Delta F(\bm X)
\end{equation}
for a duration $s$.
From \eqref{gen_functional-spacetime} it is clear that $\mathcal{G}_{\mathbb{A}}\mathbf{1}_\mathbb{A}\le 0$ and returning to \eqref{rho_prop_3.2_infgen}, we see that almost-invariant sets $\mathbb{A}$ should have $\mathcal{G}_{\mathbb{A}}\mathbf{1}_\mathbb{A}\lessapprox 0$.
To obtain multiple almost-invariant sets $A$, following \cite{FJK13} (see also \cite{FK17,FKS20}), one computes several leading real eigenfunctions of the generator $G$, where
 \begin{equation}
     \label{gen_pf_limit}
     \mathcal{G}F := \lim\limits_{s \rightarrow 0} \frac{\mathcal{P}^s F - F}{s}.
 \end{equation}
In practice, we will compute eigenfunctions of the $L^2$-adjoint of this operator; see \cite{BF24} for details.
That is, we solve 
\begin{align}
    \label{eigprob_Inflated}
\mathcal{G^*}F^*&=\Lambda F^*\,,\nonumber\\ 
\nabla F(\bm x)\cdot \bm N(\bm x)&=0\;\mbox{ for }\;\bm X\in\partial \mathbb{M},
\end{align}
where
\begin{equation}
\label{gen_inflated_adj}
\mathcal{G}^*F(t,\bm x)=\bm U(t,\bm x)\cdot\nabla F(t,\bm x)+\frac{\bm \Sigma^2}{2}\Delta F(t,\bm x).
\end{equation}
We have periodic boundary in the $x$ and $y$ coordinates, and apply Neumann boundary conditions on $\partial \mathbb{M}=(\{0,\tau\}\times M)\cup ([0,\tau]\times \partial M)$,
where $\partial M$ consists of the floor and ceiling of our three-dimensional physical domain.

The spectrum of $\mathcal{G}^*$ (and of $\mathcal{G}$) consists of eigenvalues $\Lambda_k, k=1,2,3,\ldots$ with non-positive real part, ordered by decreasing real part.
The eigenvalue $\Lambda_1=0$ has corresponding eigenfunction $F^*_1\equiv 1$.
We separate the real eigenvalues into two classes \cite{BF24,FK23}. Firstly, so-called \textit{temporal eigenfunctions}  with no dependence on $\bm{x}$;  they have the form $F_k^*(t,\bm x)=\cos(k\pi t/\tau)$, with corresponding eigenvalues $-(ak\pi/\tau)^2/2$.
Secondly, the remainder of the real eigenfunctions we call \textit{spatial eigenfunctions};  it is these that encode the quasi-stationary almost-invariant families.

\subsection{Numerical estimation of the inflated generator} \label{sec:InfGen_NumEst}

% We now outline the computational steps necessary to numerically implement the inflated generator method on a flow system that is three-dimensional in space. Code for the numerical inflated generator method is available on GitHub (\aleks{details to follow}) and we direct the reader to Appendix A for details on how to improve the computational efficiency of this method.\gf{Suggest putting in main text instead, much more briefly.}
%when applying it to three-dimensional velocity systems which will involve much more computationally expensive processes; such as multiplication between very large matrices of dimension $\sim 10^{6} \times 10^{6}$.

\begin{figure}[htb]
      \centering
\includegraphics[width=0.45\textwidth]{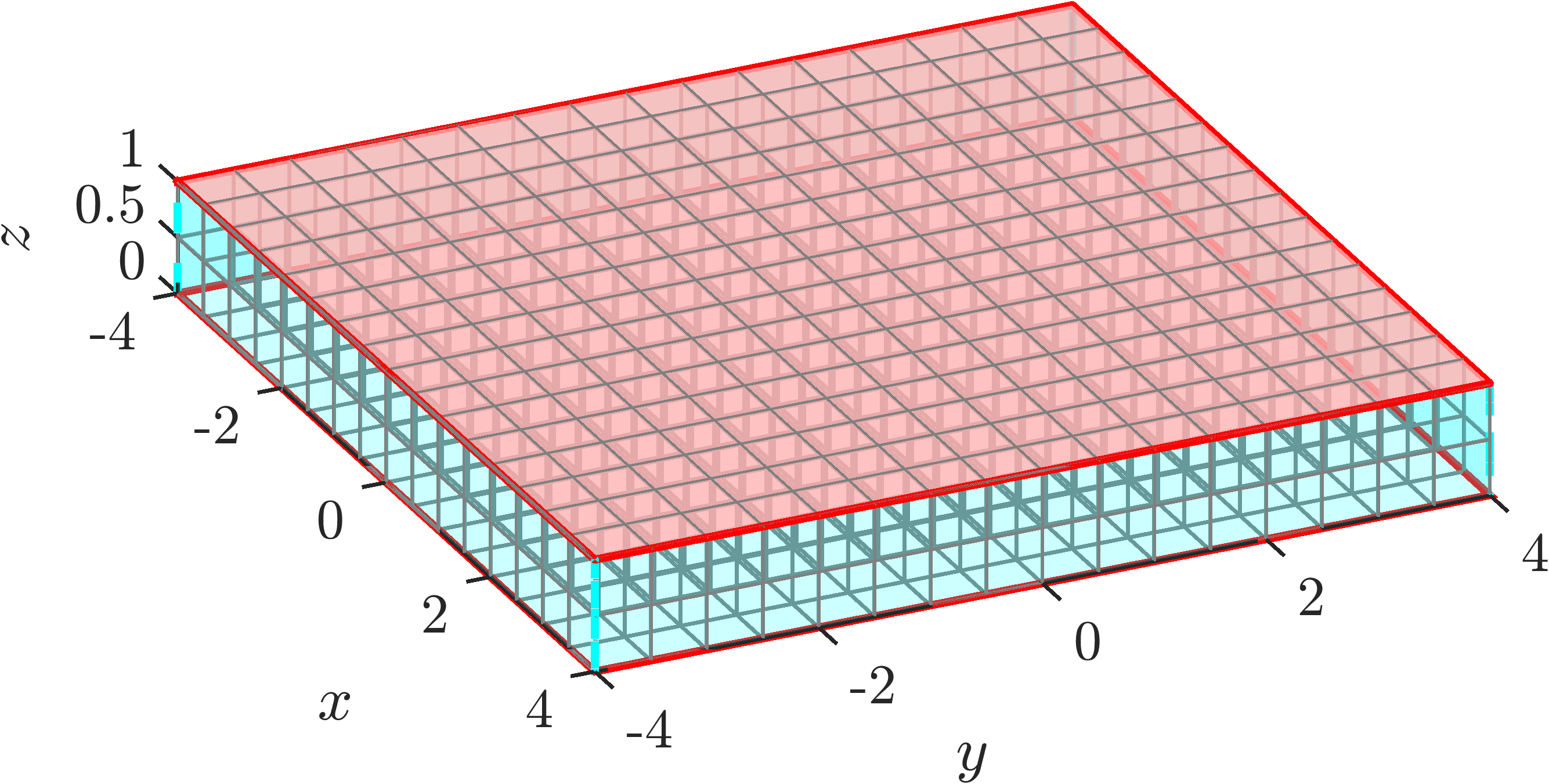}
      \caption{Sketch of the spatial domain $M$ divided into cubic cells for the construction of the spatial generators $G_t$ on each time slice with side length $\ell$. We use periodic boundary conditions on the side walls of the domain (which are highlighted in cyan) and Dirichlet boundary conditions for the temperature field at the top and bottom planes (which are highlighted in red).}
      \label{fig:spat_dom_cubes}
\end{figure}

We discretise our spacetime domain $\mathbb{M} = [0,\tau] \times M$, where  $M = [-4,4) \times [-4,4) \times [0,1] \subset\mathbb{R}^3$, by splitting the time interval $[0,\tau]$ into equidistant time steps $0 = t_0 < t_1 < \cdots < t_{K-1} < t_K =  \tau$, each separated by $\Delta t = \tau/K$, and subdividing $M$ into $N$ cubes $B_1,\ldots,B_N$ of equal side length $\ell$. Figure \ref{fig:spat_dom_cubes} shows an example of what $M$ looks like after it has been split into $N$ cubes. %Essentially, $\mathbb{M}$ is discretised into $K+1$ copies of $M$ with one copy for each time slice, in the fashion $\mathbb{M} = \cup_{k = 0}^{K} \{t_k\} \times M $, with the velocity on the $k$-th copy of $M$ being $\bm u(t_k,\bm x) = \bm u(t_k,x,y,z)$.

\subsubsection{Discretisation of spatial generators at each time instant}
At time $t$, we denote the operator corresponding to the right-hand-side of \eqref{fokker-planck} with $\bm{u}(\bm{x})=\bm{u}(t,\bm{x})$ by $G^t$.
To construct an estimate of $G^t$ -- called the spatial generator at time $t$ -- we  use ``Ulam's method for the generator" as introduced in \cite{FJK13}. 
This approach is related to an upwind scheme and involves computing integrals representing the rate of flux out of all of the faces of the boxes $\{B_i\}_{i=1}^N$.
%connecting box $B_i$ with box $B_j$, which we can denote as $B_i \cap B_j$, against the area of each face $A_{ij}$ (as area is two-dimensional volume, and $B_i \cap B_j$ will always be a two-dimensional square plane in this case).
Let $\bm{n}_{ij}$ denote the  unit normal vector to the face $B_i\cap B_j$ pointing out of $B_i$. 
We construct an $N\times N$ matrix $\mathbf{G}^t$ as 
\begin{equation}
    \label{spat_gen_eq}
    G^{t}_{ij}=\begin{cases}
    G^t_{{\rm drift},ij} + \dfrac{\epsilon^2}{2\ell^2},&\quad i\neq j,\\
    -\sum_{j\neq i}G^{t_k}_{ij},&\quad\mbox{otherwise,}
\end{cases}
\end{equation}
where $S$ denotes surface area and
\begin{equation}
    \label{flux_int}
     G_{{\rm drift},ij}^t=\frac{1}{V(B_j)}\int_{B_i \cap B_j}(\max\{\bm u(t,\bm x)\cdot \bm n_{ij},0\})\ d S(\bm x)
\end{equation}
denotes the deterministic (drift) component of the generator matrix $G^t$, which corresponds to the first term on the right-hand-side of \eqref{fokker-planck}. 
Note that at most six flux integrals are computed for each three-dimensional box.
%with a maximum of five integrals computed for the boxes which touch the ceiling and the floor as Neumann boundary conditions are used in the vertical dimension $z$. 
These integrals are computed numerically using the Julia package \verb"HCubature.jl"  \cite{hcubature,Genz1980}. All of the default choices for the input arguments are maintained, with the exceptions of the relative and absolute tolerances which are both set to $10^{-2}$ for the sake of computational efficiency.  %\gf{Aleks, please add order, tolerances, etc... that we used.} 
We add a small spatial diffusion when the flux integral is nonzero, using %We use these flux integrals to construct the discretised spatial generators $\mathbf{G}^{t_k}$ at each discrete time $t_k$, $k=1,\ldots,K$, with each one of these being a sparse $N \times N$ matrix where %\gf{first equation below needs line break now we have two columns. I am not sure how to do that nicely.}
%To each flux integral (so long as the flux integral is non-negative and the boxes $B_i$ and $B_j$ share a face) we add a spatial diffusion term $\epsilon^2/2\ell^2$, where $\ell$ is the side length of each box and we set 
\begin{equation}
    \label{epsilon_val}
\epsilon=\sqrt{0.1\overline{u}\ell}
\end{equation}
where $\overline{u}$ is the median RBC velocity magnitude over the entire spacetime domain $\mathbb{M}$ to ensure that the eigenvalue $\Lambda_1=0$ is simple.
The expression \eqref{epsilon_val} for $\epsilon$ is obtained by setting the term $\epsilon^2/2\ell^2$ in  \eqref{spat_gen_eq} equal to $0.05\overline{u}/\ell$, which corresponds to adding a drift of 5\% to the first expression in \eqref{spat_gen_eq};  see also  \cite{BF24} for a discussion.
%, however in short, by rearranging \eqref{epsilon_val} we see that
%\begin{equation*}
%    \epsilon^2 / 2 \ell^2 = 0.05 \overline{\mathbf{u}} / \ell,
%\end{equation*}
%hence by choosing $\epsilon$ in this fashion the level of spatial diffusion added to the flux integrals in \eqref{spat_gen_eq} is 5\% of the median RBC velocity across our spacetime domain $\mathbb{M}$ divided by $\ell$ to maintain the correct units for this diffusion.

\subsubsection{Discretization of the inflated generator}

 To obtain the discretised inflated generator, we note that \eqref{gen_inflated_adj} can be re-written as
\begin{equation}
    \label{gen_inflated_rewrite}
\mathcal{G}F(t,\bm x)=G_tF(t,\bm x)+\frac{a^2}{2}\partial_t^2F(t,\bm x).
\end{equation}
To create a discretised spacetime estimate of $\mathcal{G}$, we index first by space, then by time. 
For the terms $G^t$ in \eqref{gen_inflated_rewrite}, this means 
%On the diagonal, we insert negative sums of the outward fluxes across each row of $\mathbf{G}^{t_k}_{ij}$ so that all of the row sums equate to zero and we ensure that the left eigenvectors of $\mathbf{G}^{t_k}_{ij}$ (and hence the right eigenvectors of its $L^2$-adjoint) accurately represent particle densities for each box \cite{FJK13}. 
creating a block-diagonal spacetime (st) matrix with generator matrices $\mathbf{G}^t$ on the diagonal:
$$\mathbf{G}_{\mathrm{st}}:=\begin{pmatrix}
    \mathbf{G}^{t_0} & 0 & 0&  0 &\cdots & 0  \\
    0&\mathbf{G}^{t_1}& 0 &  0 &\cdots & 0\\
    0 & 0 & \mathbf{G}^{t_2} & 0  & & 0 \\
    0 &0  & 0& \ddots&  &\vdots & \\
        \vdots& & & \ddots&  &\vdots & \\
    0& 0 & \cdots & 0 & 0 &\mathbf{G}^{t_K}
\end{pmatrix}.$$
To construct the second term on the right-hand side of \eqref{gen_inflated_rewrite}, we use standard central differencing to estimate the Laplacian in time over the full time interval $[0,\tau]$, and pad with $N\times N$ identity matrices to respect the spacetime indexing: 
{\small
$$\mathbf{L}_{\mathrm{st}}:=\begin{pmatrix}
    -\bm I_N & \bm I_N & 0&  0 &\cdots & 0  \\
    \bm I_N&-2 \bm I_N& \bm I_N &  0 &\cdots & 0\\
    0 & \bm I_N & -2 \bm I_N & \bm I_N  & & \vdots \\
    0 &0  & \bm I_N & \ddots& \ddots &0  \\
        \vdots&\vdots & & \ddots& -2\bm I_N &\bm I_N  \\
    0& 0 & \cdots & 0 & \bm I_N &- \bm I_N
\end{pmatrix}/\Delta t^2.
$$}
The discretisation of $\mathcal{G}$ is then
%version of \eqref{gen_inflated_rewrite} by combining the aforementioned matrices $\mathbf{G}_{\mathrm{spacetime}}$ and $\mathbf{L}_{\mathrm{spacetime}}$ in the fashion
\begin{equation}
    \label{gen_inflated_disc}
\mathbf{G}_a = \mathbf{G}_{\mathrm{st}} + (a^2/2)\mathbf{L}_{\mathrm{st}}.
\end{equation}

\paragraph{Selecting the temporal diffusion strength $a$}
%This matrix is then scaled by a factor of $a^2/2$, where $a$ is the temporal diffusion parameter and 
The temporal diffusion matrix in \eqref{gen_inflated_disc} is scaled by a temporal diffusion strength $a$.
We discuss \emph{a posteriori} and \emph{a priori} heuristics for initial choices of $a$.
Suppose that we seek approximately $Q$ quasi-stationary families and that the average lifetimes of these families are approximately $(1/\mathcal{T})\tau$.
Then $a$ should be chosen so that the $Q^{\rm th}$ spatial eigenvalue of $\mathbf{G}_a$ is approximately equal to the $\mathcal{T}^{\rm th}$ temporal eigenvalue of $\mathbf{G}_a$.
This involves an explicit computation of the eigenvalues of $\mathbf{G}_a$.

In order to create an \emph{a priori} initial estimate of $a$, we additionally require information about the domain $\mathbb{M}$ in order to estimate the temporal and non-temporal eigenvalues.
As discussed earlier, the $k^{\rm th}$ temporal eigenvalue of $\mathbf{G}_a$ is $-(ak\pi/\tau)^2/2$.
We now need to compute the eigenvalues of $(\epsilon^2/2)\Delta$ on our spatial domain $M=8S^1\times 8S^1\times [0,1]$.
The eigenfunctions will be combinations of real Fourier modes and of the form:
\begin{eqnarray*}
    \label{fourierforms}
    &&\cos(2\pi k x/8)\cos(2\pi l y/8)\cos(\pi m z)\\
    &&\cos(2\pi k x/8)\sin(2\pi l y/8)\cos(\pi m z)\\
    &&\sin(2\pi k x/8)\cos(2\pi l y/8)\cos(\pi m z)\\
    &&\sin(2\pi k x/8)\sin(2\pi l y/8)\cos(\pi m z),\\
    \end{eqnarray*}
where $k,l,m\in\mathbb{N}$, due to periodicity in the $x$ and $y$ directions and Dirichlet boundary conditions at the domain floor and ceiling.
One may verify that the leading eigenvalues of $\Delta$ on $M$ are (multiplicities in parenthesis):
$0\ (1), -\pi^2/16\ (4), -\pi^2/8\ (4), -\pi^2/4\ (4), -5\pi^2/16\ (8),$ $-\pi^2/2\ (4), -9\pi^2/16\ (4), -5\pi^2/8\ (8),\ldots$
In our experiments we select $Q=30$ and $\mathcal{T}=2$.
From the list above, we see that the $30^{\rm th}$ eigenvalue of $(\epsilon^2/2)\Delta$ is $-5\epsilon^2\pi^2/16$ and the $2^{\rm nd}$ temporal eigenvalue is $-4a^2\pi^2/2\tau^2$.
Following \S B.1 \cite{BF24}, there is numerical diffusion introduced by the Ulam-for-the-generator scheme, which means that when determining a heuristic $a$ value, we use an ``effective'' $\epsilon$ defined by 
\begin{equation}
    \label{epseff}
\epsilon_{\rm eff}=\sqrt{1.1\overline{u}\ell}\approx 0.1228.
\end{equation}
Equating these eigenvalues yields 
\begin{equation}
    \label{Q30T2}
a_{\rm init}=\sqrt{5/32}\, \epsilon_{\rm{eff}}\tau \approx 4.9532 .
\end{equation}
% \aleksadd{The heuristic $a_{\mathrm{init}}$ above was calculated for our RBC flow system given $\overline{u} \approx 0.2195$ over the full spacetime domain $\mathbb{M} = [2001,2103] \times M$, $M = [-4,4) \times [-4,4) \times [0,1]$; $\ell = 0.0625$, $\epsilon_{\mathrm{eff}} \approx 0.1228$ and $\tau = 102 \ T_f$.}
%\gf{General question to all authors: do we want to make an explicit expression here for the spatial eigenvalues so that later we can mention the value of the 30th eigenvalue in the example section later?}\js{Yes, the more detailed, we explain this to the physics readership the better.}
The drift term in \eqref{gen_inflated_rewrite} may cause many of the leading $Q=30$ non-temporal eigenvalues to be complex.
Therefore the estimate for $a_{\rm init}$ in \eqref{Q30T2}, which assumes all real eigenvalues, may need to be increased in order to match the second temporal eigenvalue with the $30^{\rm th}$ \emph{real} non-temporal (i.e.\ spatial) eigenvalue.

We remark that when $Q=\mathcal{T}=2$ (the simplest case of two quasi-stationary families with lifetime half of the time extent $\tau$), one may compute the \emph{a priori} heuristic choice for the parameter $a$ is: 
\begin{equation}
\label{a_heuristic}
a_{\mathrm{init}}=2\tau\sqrt{1.1\overline{u}\ell}/L_{\rm max}{=2\epsilon_{\rm{eff}}\tau/L_{\rm{max}}}. %\aleksadd{\approx 3.1327},
\end{equation}
%\gf{Aleks to add numerical value above} \aleks{Value added.}
%where %\aleksadd{$\epsilon_{\mathrm{eff}}$ (and by extension $\overline{u}$ and $\ell$) and $\tau$ are as listed above and} 
$L_{\rm max}$ is the longest side length of the spatial domain $M$, which in this case will be $\Gamma = 8$. Note that the units of $a_{\rm init}$ in \eqref{a_heuristic} are the same as those in \eqref{Q30T2} because $L_{\rm max}$ is already part of the numerical coefficient in \eqref{Q30T2}. 
The heuristic for $a$ in \eqref{a_heuristic} was derived in \cite{BF24}, and has been adjusted by a factor of $2^2=4$ here to account for the periodic boundary conditions in the $x$ and $y$ coordinates (see similar periodic/nonperiodic domain discussions in \cite{FK23}).

% It must be stressed that \eqref{a_heuristic} is only an initial guess for the temporal diffusion parameter $a$, and that this value will need to be readjusted after an initial calculation of the leading eigenvalues of $\mathbf{G}_a^*$ to obtain a better visualisation of the key quasi-stationary almost-invariant features of the RBC flow system. In \cite{BF24}, the value of $a$ was fine tuned in order to allow the leading spatial and temporal eigenvalues of $\mathbf{G}_a^*$ to be approximately equal. This is a good value of $a$ to start with when using the inflated generator, and if one or no more than two quasi-stationary families of almost-invariant sets are thought to exist within a flow system, with the lifespans of these families of sets roughly covering the length of the time interval considered for the inflated generator calculations, this is more than sufficient. However, if one seeks several quasi-stationary families of almost-invariant sets which exist over varying lengths of time within the spacetime domain $\mathbb{M}$, including families of sets which also may have already formed before the start of our nominated time interval or may outlive the endpoint of that interval; it may be worth increasing $a$ to allow the leading temporal eigenvalue to be preceded by several real-valued spatial eigenvalues.
%\gf{The heuristic should be made more precise based on desired spatial scales and temporal lifetimes. I still have some more to write on this. Then I am finished with section III.}

\subsection{Extracting individual quasi-stationary families from eigenvectors of the inflated generator and SEBA} \label{sec:SEBA_Alg}

We eigensolve the discretised inflated generator $\mathbf{G}_a^*$ with the Arnoldi method using the Julia package \verb"ArnoldiMethod.jl" \cite{arnoldimethod} 
%\gf{is citation correct? I did not find the attribution to the two people you cite. Perhaps just cite the package?} \aleks{The authors of the package have requested that it be cited like this, as can be seen on the GitHub page for the package: ``https://github.com/JuliaLinearAlgebra/ArnoldiMethod.jl"; either in the CITATION.cff file, or by clicking on ``Cite this repository" and clicking on the BibTeX tab (I added the version number v0.4.0 for the package myself, because the recommended BibTeX citation does not include it for some reason).} 
to obtain the leading real-valued spatial eigenvalues  $\Lambda_k^{\rm spat} < 0$ (with values closer to 0) and their corresponding eigenvectors $\mathbf{F}_k^{\rm spat}$. 
To distinguish the real-valued eigenvalues/eigenvectors (spatial and temporal) from their complex-valued counterparts, check that the imagninary part of $\Lambda_k$ is 0 within a certain tolerance ($10^{-12}$ is sufficient for this study). To distinguish the spatial eigenvalues/eigenvectors from the temporal ones, calculate the variance of the eigenvector $\mathbf{F}_k^*$ on the $K + 1$ blocks of indices $kN+1,\ldots,(k+1)N$, $k=0,\ldots,K$ (in other words, calculate the variance of the eigenvector data across each point in our spatial domain $M$ for each of the $K + 1$ time steps), then take the average of these variances. If this average spatial variance is close to 0 (again, within a specified tolerance, $10^{-8}$ was sufficient in this case), then the eigenvector is constant on all points in space for each time step and the eigenvalue/eigenvector pair is therefore temporal. Otherwise, we have a real-valued spatial eigenvalue $\Lambda_k^{\rm spat}$ and its corresponding eigenvector $\mathbf{F}_k^{\rm spat}$, the latter of which we use in the SEBA algorithm.
%;  these eigenvectors $\mathbf{V}_k^{\rm spat}$ contain signatures of the quasi-stationary families of almost-invariant sets.
%present within the spacetime domain $\mathbb{M}$ pertinent to the RBC flow system.

\begin{comment}
    Each of the leading real-valued spatial eigenvectors of the inflated generator $\mathbf{V}_k^{\rm spat}$ contains signatures of potentially multiple and not necessarily disjoint quasi-stationary families of almost-invariant sets present within a flow system. As signatures of multiple families of sets can be present within a single eigenvector, isolating single families of sets can often be difficult. \gf{We may also cite prior work with Joerg regarding the use of SEBA in these RBC systems;  these prior work already raise similar problems and contain Figures of eigenvectors with multiple coherent sets encoded.} To assist with isolating individual and disjoint families of sets from a single vector, we can rework a subset of spatial eigenvectors into an alternative basis of sparse vectors using the Sparse Eigenbasis Approximation (SEBA) algorithm \cite{FRS19} \gf{make sentence more precise.}. Each of the vectors produced from this algorithm will isolate spatiotemporal signatures of individual fluid cells \gf{check meaning of ``fluid cell'' and whether appropriate here} present within the RBC flow, and should a SEBA vector contain traces of more than one of these fluid cells, each of these will be disjoint\gf{sentence is jumbled and incorrect.}
\end{comment}
Each of the leading real-valued spatial eigenvectors of the inflated generator $\mathbf{F}_k^{\rm spat}$ contains signatures of possibly multiple quasi-stationary families of almost-invariant sets. 

To  isolate individual  families of sets from the span of the leading spatial eigenvectors, as has been done with isolating individual coherent sets within an RBC flow system \cite{Klunker20,Vieweg24}, we use the Sparse Eigenbasis Approximation (SEBA) algorithm \cite{FRS19} to derive an sparse basis for this eigenspace.
Each SEBA vector will isolate one quasi-stationary family of almost-invariant sets.

Let $\mathcal{V}=[\mathbf{F}_1^{\mathrm{spat}}|\mathbf{F}_2^{\mathrm{spat}}|...|\mathbf{F}_Q^{\mathrm{spat}}]$ be a matrix whose columns are the leading $Q$ spatial eigenvectors of the inflated generator, let $\mathcal{S}$ be a matrix whose columns form a basis of sparse vectors each of the same length as the spatial eigenvectors $\mathbf{F}_k^{\mathrm{spat}}$, and let $\mu>0$ be a small sparsity penalty.
We seek a sparse array $\mathcal{S}$ whose column space is approximately the same as the column space of $\mathcal{V}$.
We achieve this through the following optimisation problem:
\begin{equation}
\label{sebaeqn}
\mathrm{argmin}_{\mathcal{S},\mathcal{R}} \|\mathcal{V}-\mathcal{S}\mathcal{R}\|_F^2+\mu\|\mathcal{S}\|_1.
\end{equation}
We solve \eqref{sebaeqn} by alternately fixing $\mathcal{R}$ and solving for $\mathcal{S}$ exactly using soft-thresholding, and fixing $\mathcal{S}$ and solving for $\mathcal{R}$ exactly using SVD.
When the algorithm converges we are guaranteed a local optimum, which in practice is frequently the global optimum.

\begin{algorithm}[Sparse Eigenbasis Approximation (SEBA)]
\label{SEBA_Alg}
\, 

\noindent \textit{Inputs:} A matrix $\mathcal{V}$ of size $N(K+1) \times Q$, whose columns are the $Q$ leading real-valued spatial eigenvectors of $\mathbf{G}_a^*$, $\mathbf{F}_k^{\mathrm{spat}}$, $k = 1,\ldots,Q$, each of length $N(K+1)$.  %\gf{Notation $P$ already used for Perron--Frobenius operator} 
%$N(K+1)$, where $K+1$ is the number of time steps we divide our temporal interval $[t_0,t_0+\tau]$ into, and $N$ is the number of cubes we divide our spatial domain $M$ into at each of these time steps.
%\gf{this is very wordy, please use the notation for the inflated generator and define $\mathcal{V}$ in math using notation for eigenvectors.}

\noindent \textit{Outputs:} An $N(K+1) \times Q$ matrix $\mathcal{S}$, whose columns are the $Q$ SEBA vectors $\mathbf{S}_1,\ldots,\mathbf{S}_Q$ %\gf{notation for SEBA vectors missing} 
which approximately span $\{ \mathbf{F}_1^{\mathrm{spat}},\ldots,\mathbf{F}_Q^{\mathrm{spat}} \}$.
%\gf{too wordy, use math notation for clarity.}

\begin{enumerate}
   \item Define a sparsity parameter $\mu = 0.99/\sqrt{P}$ (the largest possible value) and initialise the rotation matrix $\mathcal{R} = \mathbf{I}_Q$.
   \item To avoid degeneracies, apply a tiny random perturbation to any of the eigenvectors $\mathbf{V}_j^{\rm spat}$ that are constant vectors or close to constant. In practice, this is usually only required for the trivial eigenvector $\mathbf{V}_1^{\rm spat}$.
   \item Apply a soft thresholding  to the columns of $\mathcal{S}$: $\mathcal{S}_j = C_{\mu} ((\mathcal{V} \mathcal{R}^{\top})_j)/\| C_{\mu} ((\mathcal{V} \mathcal{R}^{\top})_j) \|$, $j = 1,\ldots,Q$ where $C_{\mu} (z) = \mathrm{sign}(z) \max \{ |z| - \mu,0 \}$.
   \item Update the orthogonal matrix $\mathcal{R}$:  let $USV^\top$ be the singular value decomposition of $\mathcal{S}^{\top}\mathcal{V}$, and set $\mathcal{R}=UV^\top$.
   \item Repeat the previous two steps until the matrix two-norm between the newly calculated matrix $\mathcal{R}$ and the matrix $\mathcal{R}$ calculated from the previous step is below a certain threshold (we used $10^{-12}$).
   \item Ensure that each column of $\mathcal{S}$ is predominantly nonnegative by setting $\mathcal{S}_j \rightarrow \mathrm{sign}\left(\sum\limits_{i = 1}^{p} \mathcal{S}_{ij}\right) \mathcal{S}_j$ for $j = 1,\ldots,Q$.
   \item Scale the columns of $\mathcal{S}$ so that each one has a maximum value of 1: $\mathcal{S}_j \rightarrow \mathcal{S}_j / \max\limits_{i} \mathcal{S}_{ij} $ for $j = 1,\ldots,Q$.
   \item Re-order the columns of $\mathcal{S}$ in decreasing order of their minimum value. Those columns with 0 or negative values close to zero are more reliable and appear first in the ordering.
   %so that the most ``reliable" SEBA vectors (i.e. those with minimum value closer to 0) appear first.
\end{enumerate}
\end{algorithm}

The $Q$ columns of $\mathcal{S}$ form our basis of SEBA vectors $\{\mathbf{S}_1,\ldots,\mathbf{S}_Q\}$.

We now summarise the full numerical approach, following Algorithm 1 in \cite{BF24}, to extracting quasi-stationary families of almost-invariant sets.

\begin{algorithm}[Identify quasi-stationary families of almost-invariant sets]
\label{InfGen_Alg}
\, 

\noindent \textit{Inputs:} A state space $M$, a time-varying velocity field $\mathbf{u}$ on $M$, a time duration $[t_0,t_0 + \tau]$.

\noindent \textit{Outputs:} Estimates of families of quasi-stationary almost-invariant sets at discrete time points in $[t_0,t_0 + \tau]$.
\begin{enumerate}
   \item Discretise the spatial domain $M$ into  cubes $B_1,\ldots,B_N$, and the time interval $[t_0,t_0 + \tau]$ into time nodes $t_{k}$, $k=0,\ldots,K$.
   \item Calculate $\epsilon$ using equation \eqref{epsilon_val} and compute the discrete generator matrices $\mathbf{G}^{t_k}$ for $k=0,\ldots,K$ as described in Section \ref{sec:InfGen_NumEst} using \eqref{spat_gen_eq}.
   \item Combine the $\mathbf{G}^{t_k}$ to form $\mathbf{G}_a$ as described in Section \ref{sec:InfGen_NumEst} using \eqref{gen_inflated_disc}, where the temporal strength parameter $a$ is initially selected using \eqref{Q30T2}.
   \item Compute the eigenvalues $0=\Lambda_1>\Lambda_2\ge \cdots$ of the sparse matrix $\mathbf{G}_a^*$ whose real parts are closest to 0, along with their corresponding spacetime eigenvectors $\mathbf{F}_1^*,\mathbf{F}_2^*,\ldots$.  Note that $\mathbf{F}_1^*$ should be a constant vector $\mathbf{1}$.
   \item Remove the complex-valued eigenvectors as well as the constant-in-space temporal eigenvectors using the steps outlined at the beginning of this subsection. 
   %by inspection or numerical interrogation \gf{refer to earlier text that Aleks will add on how this is done}; 
   %\gf{Remaing text can form park of the earlier description instead of appearing here} the discretised temporal eigenvectors $\mathbf{F}_k^*$ will be constant on blocks of indices $kN+1,\ldots,(k+1)N$, $k=0,\ldots,K$.
   \item Isolate $Q$ quasi-stationary families of almost-invariant sets by applying SEBA to the leading $Q$ \textit{spatial} eigenvectors (including the trivial constant vector $\mathbf{F}_1^*$) to obtain spacetime SEBA vectors $\mathbf{S}_1,\ldots,\mathbf{S}_{Q}$. Each SEBA vector should support one of these quasi-stationary families of almost-invariant sets in spacetime.
\end{enumerate}
\end{algorithm}

\section{Results}

In this Section, we provide computational specifics for the inflated generator method applied to our RBC flow system, show relevant results obtained from these methods, and then compare the results from this method with the convective heat transfer (CHT) results shown in Section 2 in an attempt to establish a connection between strong convective heat transfer and the lack of quasi-stationary, almost-invariant flow behaviour (or conversely, a connection between low convective heat transfer and quasi-stationary almost-invariant flow behaviour) within this system. We compare these results both qualitatively, through visual comparisons of scalar fields, and quantiatively through computation of the correlation between the maxima of the SEBA vectors produced from the inflated generator method and both instantaneous and time-averaged CHT data.

\subsection{Quasi-stationary families of almost-invariant sets from the inflated generator}

\begin{comment}
    \gf{Please briefly discuss findings at other resolutions instead of just immediately launching into a specific set of numbers.}
To execute the inflated generator method on the RBC flow system, we start by discretising our spatial domain $M = [-4,4) \times [-4,4) \times [0,1]$ into $128 \times 128 \times 16 = 262144$ cubes of equal side length $\ell = 1/16 = 0.0625$. We focus on the time interval $[t_0,t_0+\tau] = [2001,2103] \ T_f$, giving $\tau = 102$ and we take 35 equispaced time steps spaced $\Delta t = 3 \ T_f$ apart. This will produce an inflated generator matrix of size $9175040 \times 9175040$ for our spacetime domain $\mathbb{M} = [2001,2103] \times M$, so constructing this matrix (through computation of the flux integrals), solving it for its eigenvalues/eigenvectors and producing SEBA vectors from the resultant real-valued spatial eigenvectors will have to be made more efficient using the techniques described in Appendix A.
\gf{The message from this opening paragraph should not be that we have a computational problem, it should instead be robustness of the results to specific discretisation resolutions.}
\end{comment}

%\gf{It may help to break this description into the steps of the algorithm...which seems not to be referred to.}
To execute the inflated generator method on the RBC flow system, we follow the steps outlined in Algorithm \ref{InfGen_Alg}, starting by discretising our spatial domain $M = [-4,4) \times [-4,4) \times [0,1]$ into $N$ cubes of equal side length $\ell$ at each of our $K + 1$ time steps. We take 35 equispaced time steps spaced $\Delta t = 3 \ T_f$ apart on the time interval $[t_0,t_0+\tau] = [2001,2103] \ T_f$ of length $\tau = 102$. We have tried three different values of $\ell$ for the inflated generator method in this case, these being 0.25, 0.125 and 0.0625 (which, respectively, gives $N = $ 4096, 32768 and 262144). In this paper we focus on the results generated with $\ell = 0.0625$ as this provides the best spatial resolution for $M$ and allows us to identify quasi-stationary families of almost-invariant sets with greater clarity.
We compute the discrete generator matrices $\mathbf{G}^{t_k}$ on each of our 35 time steps using \eqref{spat_gen_eq}. Using \eqref{epsilon_val}, we compute the value for the spatial diffusion parameter $\epsilon\approx 0.0370$, given the box side length $\ell = 0.0625$ and the median velocity $\overline{u}$ over the entire spacetime domain $\mathbb{M}$ taking an approximate value of $0.2195$. As we have a large number of flux integrals to compute in this case (at most, $6 \times 262144 = 1572864$ as each cube will have at most six neighbouring cubes), we make this process more efficient using standard multithreading in Julia so that multiple flux integrals can be evaluated at once.
We then construct the full discretised inflated generator matrix using \eqref{gen_inflated_disc} which, given $\ell = 0.0625$, will be a relatively large matrix of size $9175040 \times 9175040$ for our spacetime domain $\mathbb{M} = [2001,2103] \times M$. Using \eqref{Q30T2}, we calculate an initial heuristic for the temporal diffusion parameter $a$, $a_{\rm init} \approx 4.9532$, given $\tau = 102$ and $L_{\max} = \Gamma = 8$. We then eigensolve the discretised inflated generator for its leading eigenvalues/eigenvectors using the Arnoldi method. To make this process more efficient for a large inflated generator matrix, large matrices such as the inflated generator and the Krylov subspace matrix for the Arnoldi method are defined on a GPU so that matrix multiplications that form part of the Arnoldi method can be executed more quickly.
We construct and eigensolve the inflated generator matrix using the initial heuristic for $a$ \eqref{Q30T2}. 

\begin{figure}[htb]
      \centering
\includegraphics[width=0.45\textwidth]{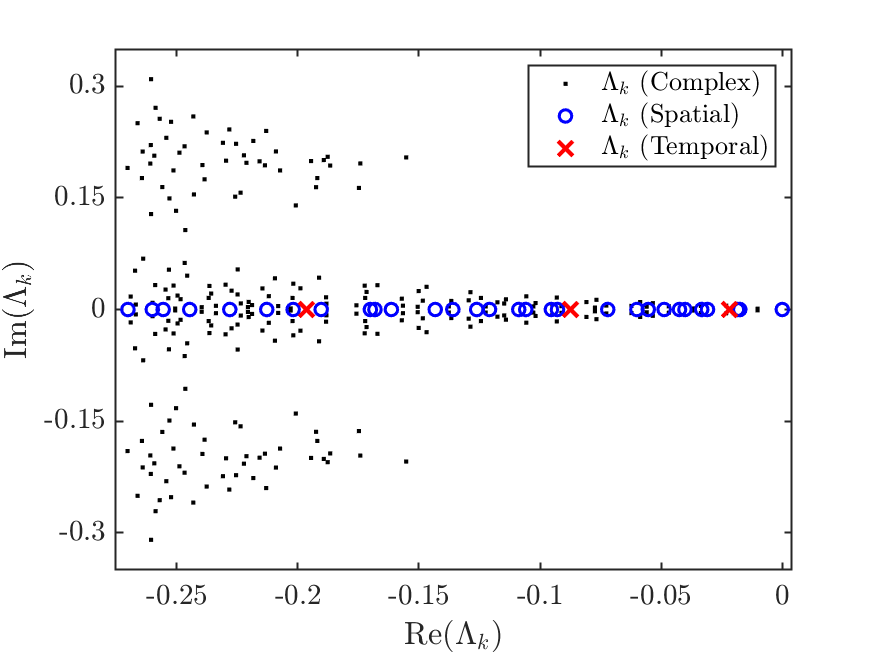}
      \caption{The spectrum of the inflated generator for the three-dimensional RBC flow with $a = a_{\mathrm{init}} \approx 4.95$, showing the leading thirty real-valued spatial eigenvalues (indicated by blue circles), the first three temporal eigenvalues (indicated by red crosses), and an extensive collection of complex valued eigenvalues (indicated by black dots).}
      \label{fig:rbc_infgen_spectrum}
\end{figure}

Figure \ref{fig:rbc_infgen_spectrum} shows the corresponding spectrum.
The second temporal eigenvalue roughly matches the thirtieth non-temporal eigenvalue, however many of the non-temporal eigenvalues with larger real part than the second temporal eigenvalue are a mix of complex and spatial eigenvalues. 
In order for the second temporal eigenvalue to match with the thirtieth \textit{spatial} eigenvalue (which in this case is the 285th overall), the value of the parameter $a$ would have to be increased to around $a=9.15$. 
However, such an increase in $a$ would reduce the amount of temporal variation between time slices of the inflated generator eigenvectors and the SEBA vectors. 
To emphasise this temporal variation, we reduce $a$ slightly to a value of 4.1 for subsequent experiments with a single value of $a$.

%-------------------------------------------
\begin{figure*}
      \centering
\includegraphics[width=\textwidth]{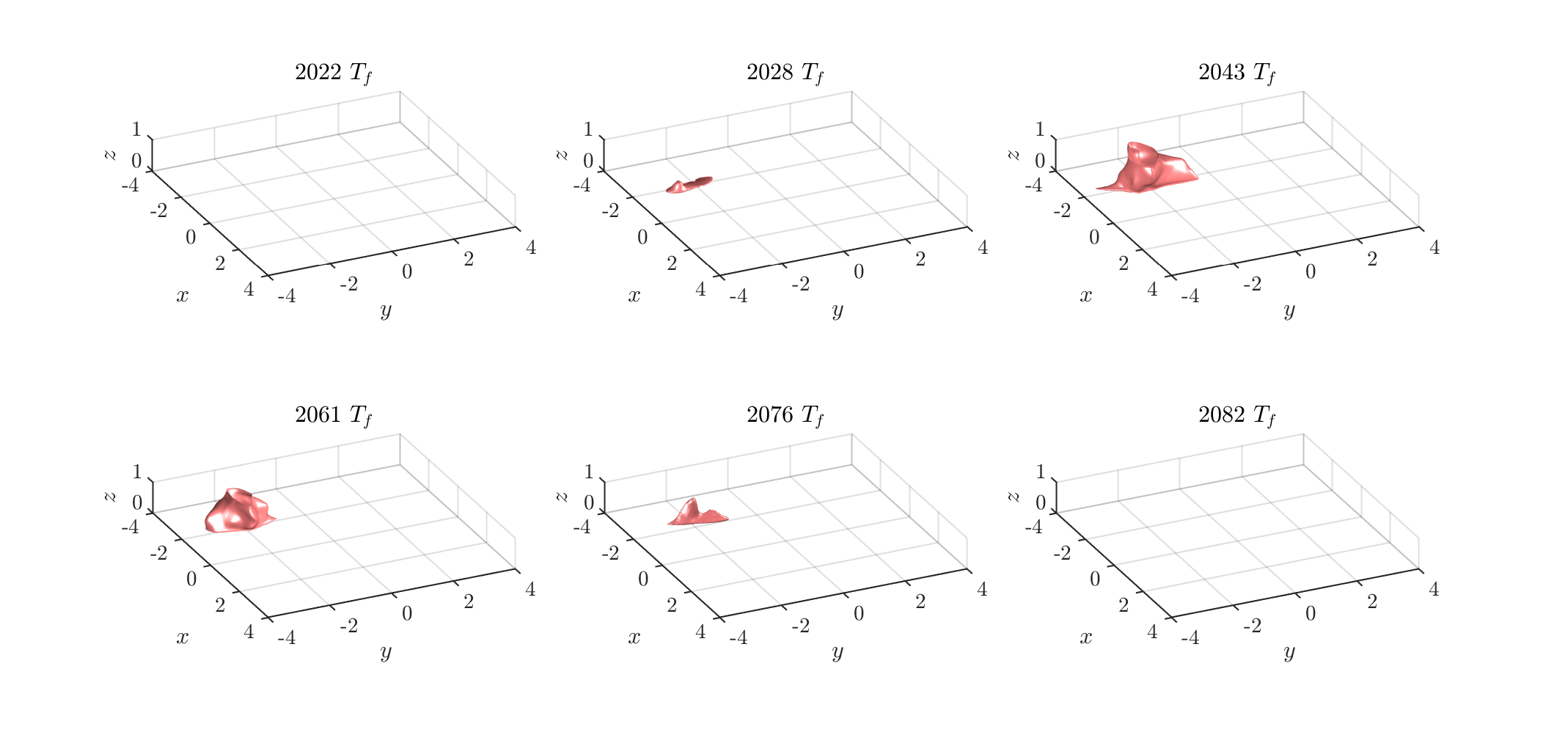}
      \caption{The evolution of a quasi-stationary family of almost-invariant sets (a transient plume of fluid forming a metastable parcel) within the RBC flow over six discrete time steps as identified through the SEBA vector $\mathbf{S}_{21}$. A cutoff of 0.4 has been applied to the SEBA vector.}
      \label{fig:rbc_seba_spacetime_example}
\end{figure*}

Since we seek several quasi-stationary families of almost-invariant sets within this flow system (in this case, roughly 20-30 within the time interval of interest), we take the corresponding eigenvectors $\mathbf{F}_1^{\mathrm{spat}},\ldots,\mathbf{F}_{30}^{\mathrm{spat}}$ of these 30 spatial eigenvalues and we insert these into the SEBA algorithm in order to isolate signatures of the quasi-stationary families of almost-invariant sets present within the RBC flow system. Since the matrices $\mathcal{V}$, $\mathcal{R}$ and $\mathcal{S}$ will all be large, in similar fashion to what was done with the Arnoldi method we define all of these matrices on a GPU to make Algorithm \ref{SEBA_Alg}, in particular steps 3-5 of this Algorithm, more efficient.
After calculating these 30 SEBA vectors $\mathbf{S}_1,\ldots,\mathbf{S}_{30}$, we further improve the spatial resolution of these spacetime vectors by interpolating them over a uniform three-dimensional grid of points within our spatial domain $M$ with mesh size $\ell/2 = 0.03125$.

Figure \ref{fig:rbc_seba_spacetime_example} shows a typical example of a quasi-stationary family of almost-invariant sets detected for the plane-layer RBC flow system identified through three-dimensional isosurfaces of the SEBA vector $\mathbf{S}_{21}$ taken for example at a level of 0.4 at discrete time steps. As can be seen in Fig. \ref{fig:rbc_seba_spacetime_example}, at time $t=2022 T_f$ our spatial domain $M$ is blank, indicating that our family of sets (which will represent a transient  plume of fluid comprising an almost-invariant parcel) has not formed yet. However, at $t = 2028 T_f$ a red object (representing a member of our family of sets) begins to emerge along the bottom of our fluid layer. This object then begins to grow in size, before reaching the top of our cell at time $t= 2043 T_f$ to form our full metastable/almost-invariant flow object. Once fully formed, this transient family of almost-invariant sets retains its shape for roughly 20-25 units of time, until it begins to shrink back down towards the bottom of our cell at around time $t\approx  2061 T_f$. By time $t=2076 T_f$, the family of sets has been reduced to a small object sitting at the bottom of the cell, and by time $t=2082 T_f$ the family of sets has vanished completely. 

\begin{figure*}
      \centering
\includegraphics[width=\textwidth]{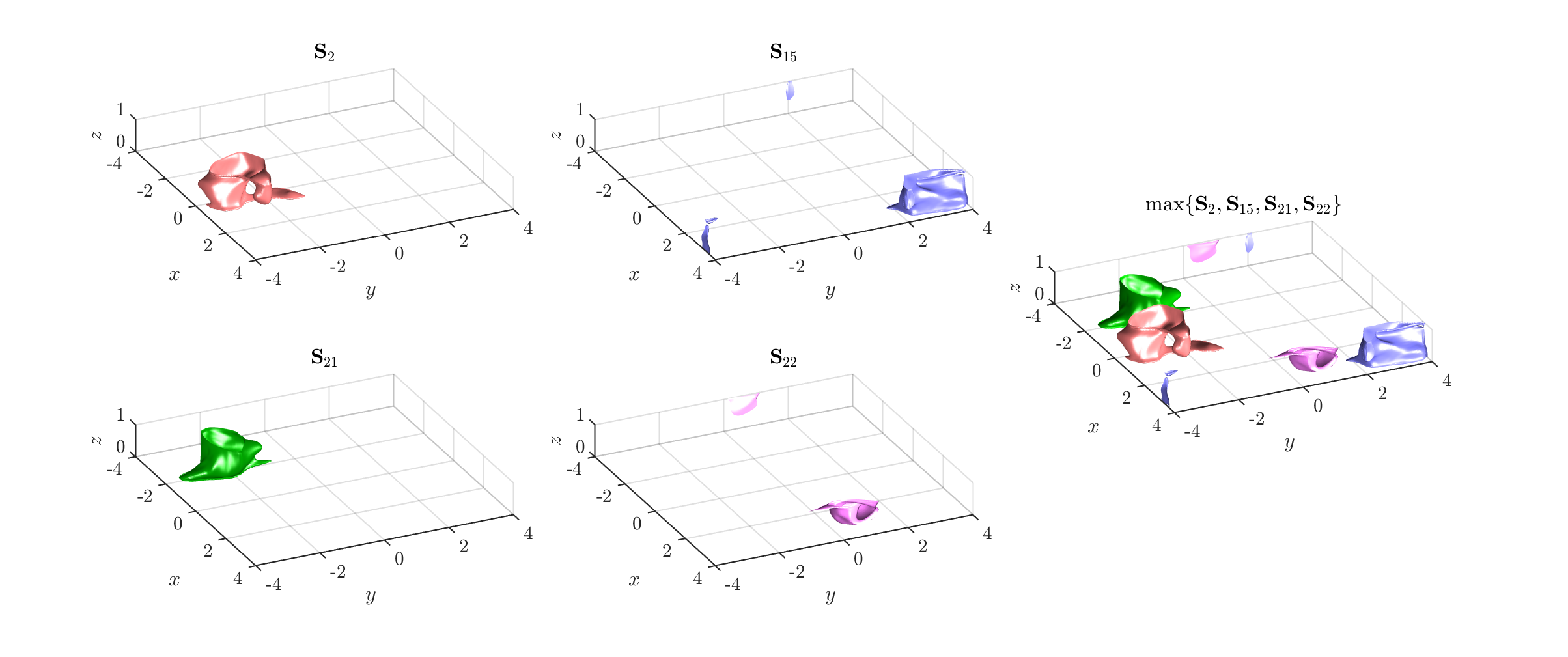}
      \caption{Three-dimensional isosurfaces for 4 of the 30 SEBA vectors computed for the RBC flow (namely $\mathbf{S}_{2}$, $\mathbf{S}_{15}$, $\mathbf{S}_{21}$ and $\mathbf{S}_{22}$) taken at the central time step 2052 $T_f$. A cutoff of 0.4 has been applied to each of these SEBA vectors. In a fifth panel (right), we superimpose these four vectors on a single three-dimensional axis to show these four families of sets co-existing within the RBC flow system at time 2052 $T_f$.}
      \label{fig:rbc_seba_individual_3D}
\end{figure*}

\begin{figure*}
      \centering
\includegraphics[width=\textwidth]{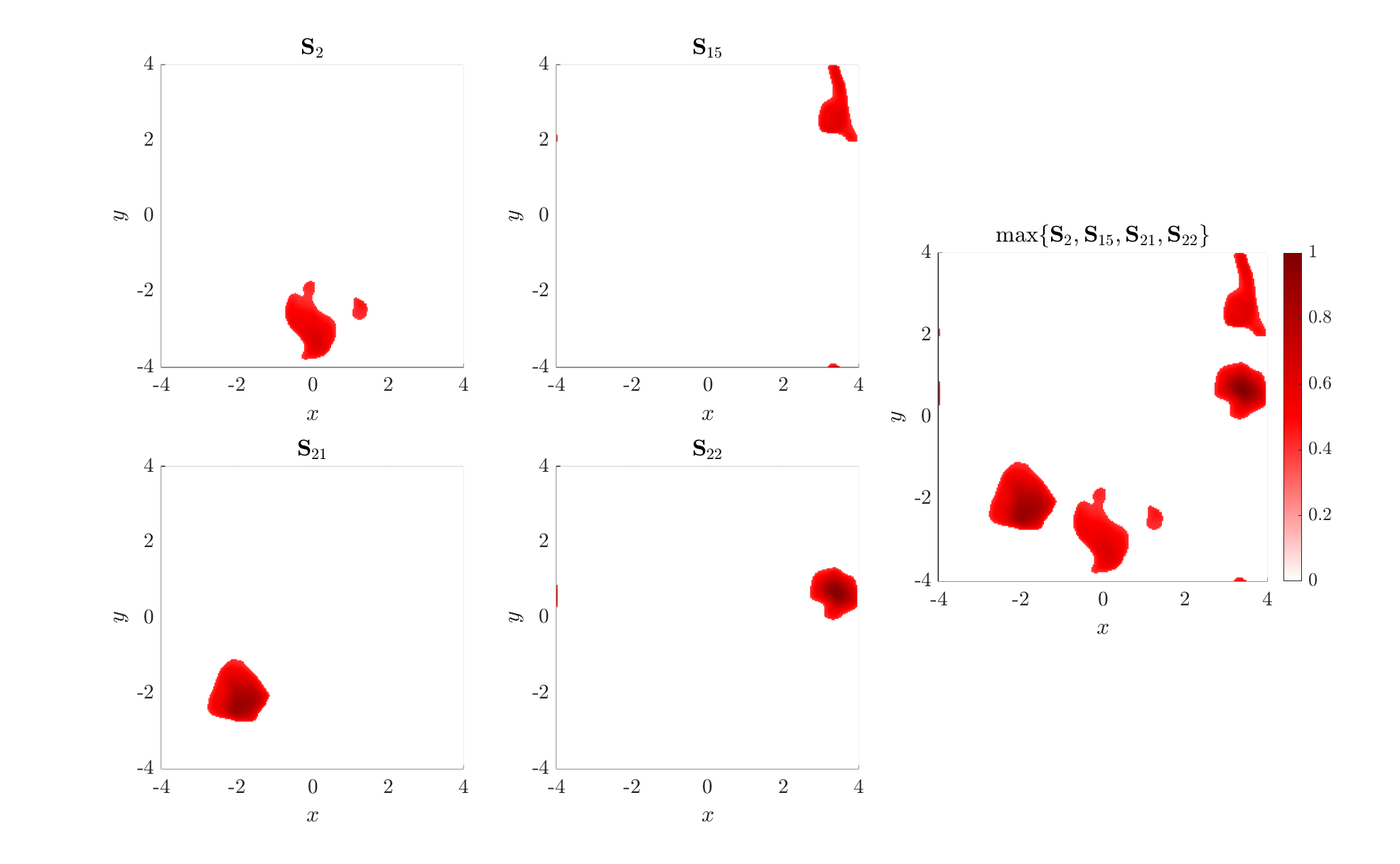}
      \caption{Snapshots of the same 4 SEBA vectors shown in Figure \ref{fig:rbc_seba_individual_3D} restricted to the horizontal $xy$-midplane (with $z = 0.5$) at the central time step 2052 $T_f$. A cutoff of 0.4 has been applied to each of these SEBA vectors. In a fifth panel (right), we take the maximum of these four vectors and restrict this to the $xy$-midplane to demonstrate the co-existence of these four families of sets found through the four SEBA vectors at time 2052 $T_f$.}
      \label{fig:rbc_seba_individual_2DMP}
\end{figure*}

Several quasi-stationary families of almost-invariant sets are present simultaneously. To illustrate this, Fig. \ref{fig:rbc_seba_individual_3D} shows four of the many distinct quasi-stationary families of almost-invariant sets present at time $t=2052 T_f$ (the central point of our time interval), found through four different SEBA vectors ($\mathbf{S}_{2}$, $\mathbf{S}_{15}$, $\mathbf{S}_{21}$ and $\mathbf{S}_{22}$). In a fifth panel within Figure \ref{fig:rbc_seba_individual_3D}, we superposition these four vectors to illustrate the four families of sets existing simultaneously at this moment in time. In Fig. \ref{fig:rbc_seba_individual_2DMP}, we restrict these four SEBA vectors, along with the superposition panel, to the $xy$-midplane with $z = 0.5$. In Figure \ref{fig:rbc_seba_individual_2DMP}, two-dimensional restrictions of quasi-stationary families of almost-invariant sets are identifiable through solid or dark red objects (corresponding to SEBA values of isolevel 0.4 or higher). We again apply a cutoff of 0.4 to both Figs \ref{fig:rbc_seba_individual_3D} and \ref{fig:rbc_seba_individual_2DMP} so that we can easily visualise the individual quasi-stationary families of almost-invariant sets.

\begin{figure*}
      \centering
\includegraphics[width=\textwidth]{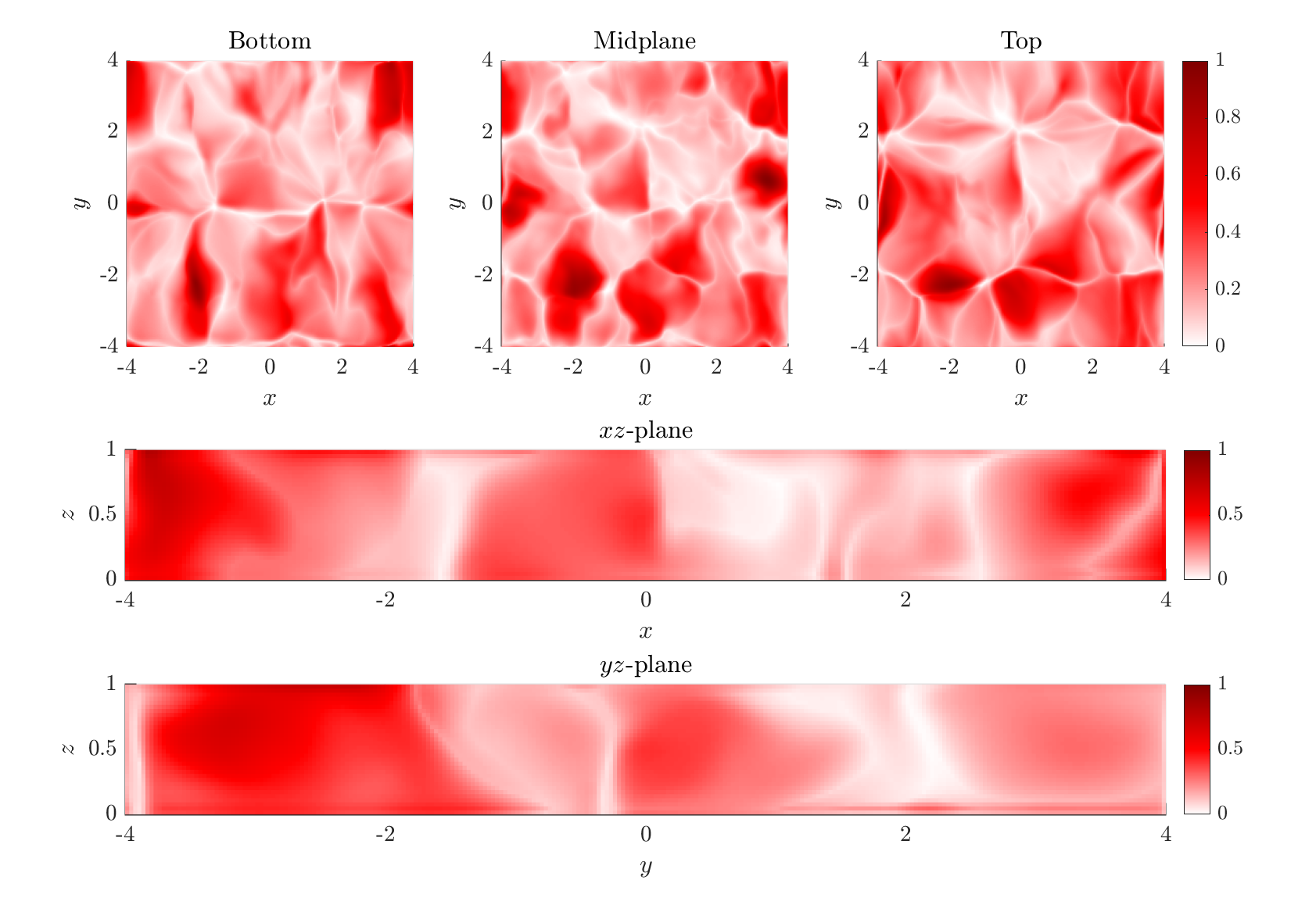}
      \caption{Snapshots of the maximum of 30 SEBA vectors computed for the RBC flow taken at 2052 $T_f$ along five two-dimensional restrictions of the spatial domain $M$. We take horizontal restrictions of the SEBA maxima (shown in the top row) along the $xy$-plane at three vertical levels (from left to right, the ``bottom" of the cell ($z = 0.0625$), the ``midplane" ($z = 0.5$) and the ``top" of the cell ($z = 0.9375$)); and we take vertical restrictions of the SEBA maxima in the $xz$-midplane with $y = 0$ (middle row) and in the $yz$-midplane with $x = 0$ (bottom row).}
      \label{fig:rbc_sebamax_2052Tf}
\end{figure*}

In order to produce one complete picture of the quasi-stationary, almost-invariant flow behaviour present across our entire spacetime domain $\mathbb{M}$, we calculate the maximum of the 30 SEBA vectors $\mathbf{S}_{\rm max}$, where
\begin{equation*}
    (\mathbf{S}_{\rm max})_i = \max\limits_{j} \mathbf{S}_{ij}
\end{equation*}
over all $j = 1,\ldots,30$. The full collection of quasi-stationary, almost-invariant plumes of fluid obtained from $\mathbf{S}_{\rm max}$ should fill most  of our spatial domain $M$ on each time slice, with the remaining filaments of space not covered by these plumes representing regions of faster and less metastable fluid transport.
After taking the maximum of the 30 spacetime SEBA vectors $\mathbf{S}_{\rm max}$ (including the four shown in Figs. \ref{fig:rbc_seba_individual_3D} and \ref{fig:rbc_seba_individual_2DMP}), we use $\mathbf{S}_{\rm max}$ to produce the images shown in Fig. \ref{fig:rbc_sebamax_2052Tf}. We plot horizontal restrictions of the SEBA maxima at three levels in the $xy$-plane, namely the ``Top" (positioned along $z = 1-\ell = 0.9375$, which is not on the ceiling of the cell but is still within the upper thermal boundary layer for the cooling plate at the top of the cell), the ``Midplane" ($z = 0.5$, the middlemost horizontal layer), and the ``Bottom" ($z = \ell = 0.0625$, which is within the lower thermal boundary layer for the heating plate at the bottom of the cell). We also restrict the SEBA maxima to two vertical midplanes, one in the $xz$-plane (with $y = 0$) and one in the $yz$-plane (with $x = 0$).

In Fig. \ref{fig:rbc_sebamax_2052Tf}, as was the case in Fig. \ref{fig:rbc_seba_individual_2DMP}, the regions of our spatial domain colored in red or deep red (corresponding to maximal SEBA values of approximately 0.4 or higher) represent quasi-stationary, almost-invariant fluid objects present within the RBC flow. Regions colored in light pink represent quasi-stationary families of almost-invariant sets which are beginning to either emerge or dissipate; while the thin white streaks or filaments correspond to parts of our domain where no almost-invariant behavior is present, and which are more likely to correspond to thin passageways through which parcels of less metastable fluid flow towards the top wall after being heated, or sink towards the bottom wall after being cooled. 

\subsection{Qualitative comparisons between SEBA and CHT}

\begin{figure*}
\centering
\includegraphics[width=0.9\textwidth]{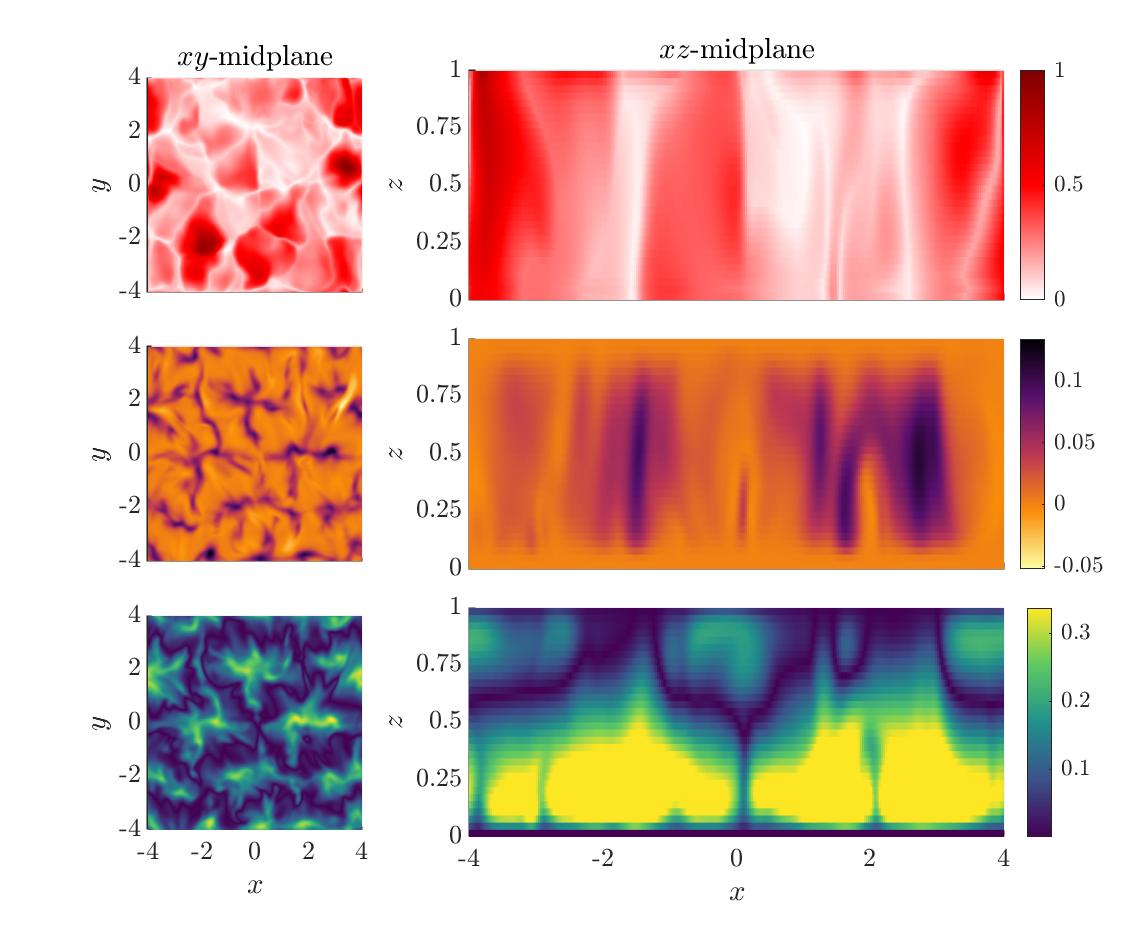}
      \caption{The maximum of 30 SEBA vectors (top row), CHT averaged over $5 \ T_f$ (middle row) and $|\theta|$ also averaged over $5 \ T_f$ (bottom row) for the RBC flow taken at time 2052 $T_f$. We restrict these three-dimensional quantities to the horizontal $xy$-midplane (with $z = 0.5$) on the left, and to the vertical $xz$-midplane (with $y = 0$) on the right.}
      \label{fig:rbc_sebaAndcht_2052Tf}
\end{figure*}

We now focus on a connection between the quasi-stationary families of almost-invariant sets identified from the SEBA vectors and the values of convective heat transfer (CHT) data for the RBC flow system. We start with a more qualitative approach to doing this by comparing the scalar fields for the SEBA vectors and the CHT data. Figure \ref{fig:rbc_sebaAndcht_2052Tf} shows the scalar fields of the maximum of 30 SEBA vectors computed from the inflated generator method with $a = 4.1$ (as discussed in the previous subsection) and the CHT data averaged over 5 $T_f$ in the horizontal $xy$-midplane and the vertical $xz$-midplane %\aleksadd{(the absolute value of temperature fluctuation $\theta$, also averaged over 5 $T_f$ is also plotted in Figure \ref{fig:rbc_sebaAndcht_2052Tf}, however we will concentrate more on comparing this with SEBA later on)}.
The full three-dimensional CHT data is averaged over a window of length 5 $T_f$, $[t_k-2,t_k+2]$ with $t_k = 2052 \ T_f$, and is then restricted to the two-dimensional horizontal $xy$-midplane and the vertical $xz$-midplane. We must stress that the SEBA spacetime vectors and CHT data is three-dimensional and defined over the entire spatial domain $M$, however, for ease of display we restrict the data to these two midplanes for this qualitative analysis. We focus on more quantitative comparisons of the full three-dimensional data in the next two subsections.

For brevity, we restrict the qualitative analysis in this paper to one value of the temporal diffusion parameter $a = 4.1$ for the inflated generator method to produce the SEBA vectors, and one length of time over which we average the CHT data (5 $T_f$). To the naked eye, some connections can be drawn between the SEBA and CHT fields in Fig.\ \ref{fig:rbc_sebaAndcht_2052Tf}. For instance, at a region in the neighbourhood of the point $(0,2)$ in the $xy$-midplane, the crucifix-shaped purple object in the CHT field roughly coincides with a web of white filaments centred at the same location in the maximum SEBA field. In the $xz$-midplane, we can also spot some of these similarities, such as a thin white colored stem roughly close to the vertical line $x = -1.5$ in the maximum SEBA field lining up with a purple vertical object (a little thicker than the white stem) in the CHT field. Other examples like these can be deduced from the scalar fields, signifying a potential connection between low SEBA (and therefore less metastable fluid behavior) and high average CHT (corresponding to faster fluid movement as heated fluid rises from the bottom of our cell and cooled fluid sinks from the top). That said, while the regions of high CHT usually take the shape of thicker purple or black blobs, the white ridges of minimal SEBA are little more than wafer-thin filaments which separate the quasi-stationary families of almost-invariant sets.

This connection is not perfect, as there are a few visual inconsistencies between the two fields. This is particularly true in the opposite case, where we seek connections between high SEBA (quasi-stationary families of almost-invariant sets) and low CHT (reduced transfer of heat from reduced fluid movement as large metastable objects form). As we can see in Fig.\ \ref{fig:rbc_sebaAndcht_2052Tf}, while there are some overlaps between the red regions of the SEBA fields and the deep orange or yellow regions of the CHT fields, the shapes of the red objects in the SEBA field do not completely line up with orange or yellow objects in the CHT field. These observations suggest a qualitative connection (albeit weak) does exist between SEBA and CHT when the movement of fluid is more rapid and convective heat transfer is stronger, but not within regions of the domain where fluid movement is more quasi-stationary and metastable (or at least, not to as great an extent).

In the $xy$-midplane, we note that the $|\theta|$ field, much like the CHT ($H$) field, does share common features with the SEBA field, particularly between the thin white streaks of the SEBA field and the yellow regions of high $|\theta|$, reinforcing that there is some connection between regions of less metastable fluid advection and increased temperature fluctuation (in absolute value, disregarding whether this is a net positive or negative fluctuation). However, when we focus on the vertical $xz$-midplane, it is much more difficult to develop some connection between SEBA and the $|\theta|$ field, in contrast to CHT.

Thus, it appears that it is difficult to detect transient almost-invariant regions in this turbulent flow using physical properties of the fluid directly, and one requires the more specialised inflated generator method, which targets precisely the emergence and dissipation of almost-invariant objects.

\begin{figure*}
      \centering
\includegraphics[width=0.95\textwidth]{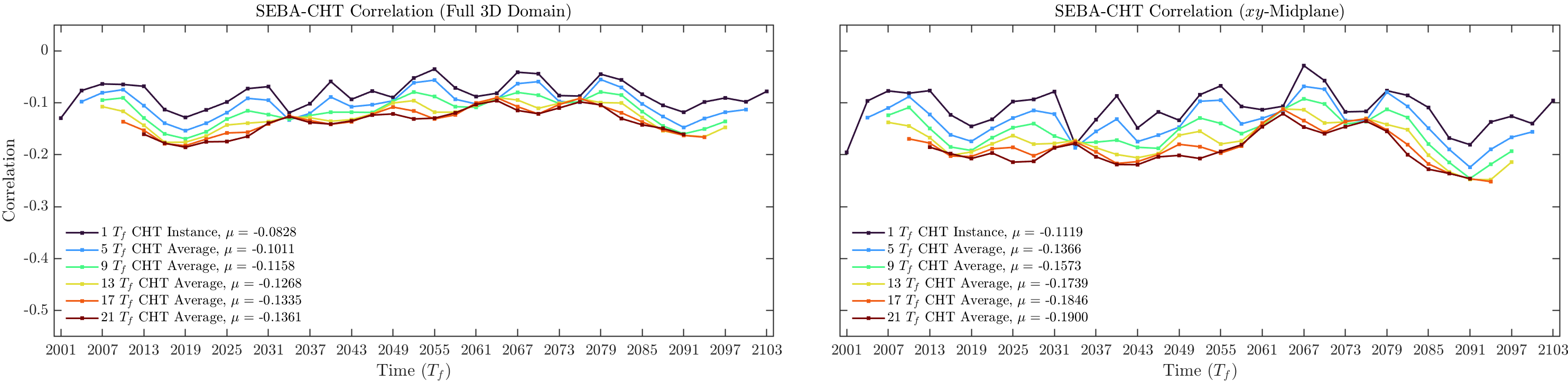}
      \caption{Plots of correlation values between SEBA and CHT against time. In the left-hand Figure, we take SEBA and CHT data over the entire three-dimensional spatial domain $M$ and in the right-hand Figure we restrict the data to the $xy$-midplane. The SEBA vectors have been computed from the inflated generator method with $a$ set to 4.1. Each colour represents the length of time over which the CHT data is averaged, and the correlation values are plotted at the midpoints $t_k$ of the time intervals $[t_k - m,t_k+m]$ over which we compute the averages for the CHT data. Included in the plot's legend are the means of the correlation values $\mu$ taken over time for each curve.}
      \label{fig:rbc_cht_corr}
\end{figure*}

\begin{figure*}
      \centering
\includegraphics[width=0.95\textwidth]{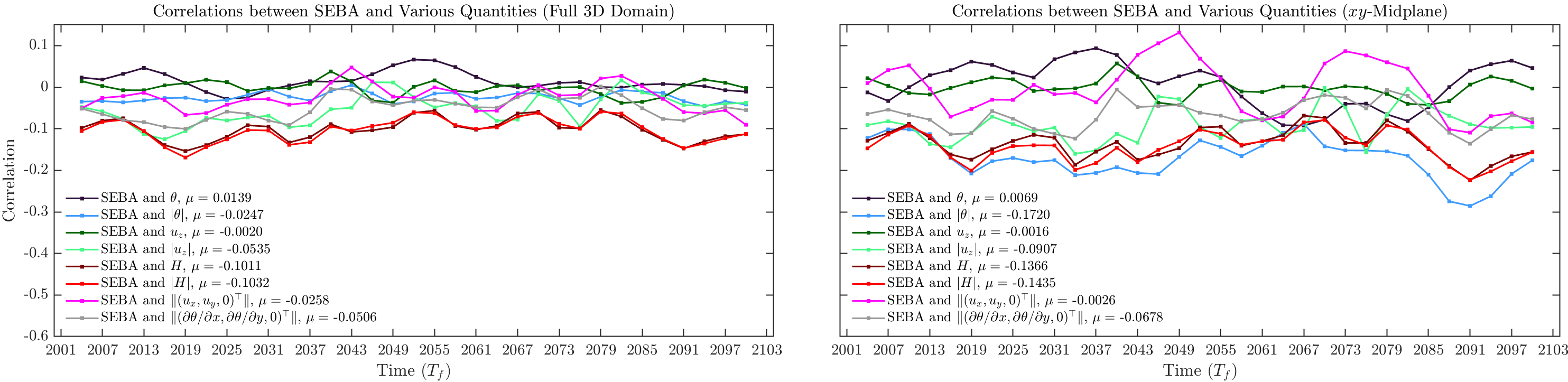}
      \caption{Plots of correlation values between SEBA and various other quantities against time: %The quantities in question are those plotted in Figure \ref{fig:rbc_cht},
      temperature fluctuation $\theta$, vertical velocity $u_z$ and convective heat transfer $H$; along with the absolute values of these quantities $|\theta|, |u_z|$, $|H|$; horizontal speed $\| (u_x,u_y,0)^\top\|$, and the norm of the horizontal gradient of $\theta$, $\| (\partial \theta / \partial x,\partial \theta / \partial y,0)^\top\|$. In the left-hand panel, we compute these correlations over the entire three-dimensional spatial domain $M$ and in the right-hand panel we restrict these calculations to the $xy$-midplane. The SEBA vectors have been computed from the inflated generator method with $a$ set to 4.1. Each colour represents one of the quantities of interest.
      %, either $\theta$, $u_z$, $H$; the absolute values of one of these quantities, the horizontal speed $\| (u_x,u_y,0)^\top\|$, or the norm of the horizontal $\theta$ gradient $\| (\partial \theta / \partial x,\partial \theta / \partial y,0)^\top\|$.
      The data for each of the quantities
      %$\theta$, $u_z$, $H$, $\| (u_x,u_y,0)^\top\|$ and $\| (\partial \theta / \partial x,\partial \theta / \partial y,0)^\top\|$ 
      is averaged over a duration of 5 $T_f$,  and the correlation values are plotted at the midpoints $t_k$ of the time intervals $[t_k - 2,t_k+2]$ over which we compute the averages for these quantities. Included in the plot's legend are the means of the correlation values $\mu$ taken over time for each curve.}
      \label{fig:rbc_cor_variousQuants}
\end{figure*}

\subsection{Correlation between SEBA and CHT}

We quantitatively compare the SEBA field and other fields such as CHT by computing correlations.
Let $\mathrm{SEBA}_{t,i}$ and $\mathrm{CHT}_{t,i}$ respectively denote the values of the SEBA and CHT fields at time $t$ and pixel $i$.
Pearson's correlation coefficient at time $t$ is computed with 
\begin{align}
\label{eq:Pearsons_r}
&\mathrm{Corr}_t(\mathrm{SEBA},\mathrm{CHT}):=\nonumber\\
&    \frac{\sum\limits_{i = 1}^{N}(\mathrm{SEBA}_{t,i} - \overline{\mathrm{SEBA}}_{t})(\mathrm{CHT}_{t,i} - \overline{\mathrm{CHT}}_{t})}{\sqrt{\sum\limits_{i = 1}^{N}(\mathrm{SEBA}_{t,i} - \overline{\mathrm{SEBA}}_{t})^2} \sqrt{\sum\limits_{i = 1}^{N}(\mathrm{CHT}_{t,i} - \overline{\mathrm{CHT}}_{t})^2}},
\end{align}
where 
\begin{align*}
\overline{\mathrm{SEBA}}_{t}=\frac{1}{N}\sum_{i=1}^N \mathrm{SEBA}_{t,i} \quad\mbox{and}\quad \overline{\mathrm{CHT}}_{t}=\frac{1}{N}\sum_{i=1}^N \mathrm{CHT}_{t,i}.
\end{align*}
In Fig. \ref{fig:rbc_cht_corr}, we plot the correlation between the full three-dimensional SEBA data and CHT data against time at every 3 $T_f$ (as this is the temporal resolution of the SEBA spacetime vectors). We also plot the correlation between SEBA and CHT restricted to the $xy$-midplane. In this instance, we focus on SEBA data computed from the inflated generator with $a = 4.1$, while the CHT data is taken at instantaneous moments in time, and is also averaged over subintervals of time of length 5, 9, 13, 17 and 21. The subintervals in question take the form $[t_k-m,t_k + m]$, where $t_k$ is the central time point of the interval and $m = 2p$, $p = 1,\ldots,5$. The correlation values computed between SEBA and average CHT are plotted at these central time points $t_k$.

From the left hand graph in Fig. \ref{fig:rbc_cht_corr}, we observe a (weak) negative correlation between SEBA and CHT. This correlation is considerably weaker when we calculate it using single-time instances of CHT, however as the length of the time window over which we compute CHT averages increases, the correlation between SEBA and CHT becomes more negative and therefore improves. This implies that the removal of noise/diffusion from the CHT data by averaging the data helps improve the correlation between CHT and SEBA. One could argue that it is possible to improve the correlation values even further by taking CHT averages over longer time windows than those already considered (e.g. in the region of 60-90 $T_f$). However, the observed ``bunching" of the curves as the time window length increases suggests that we are approaching some limit for the improvement of these correlation values, where increasing the window of time over which we average the CHT data will only be effective at improving the correlation up to a point.

\begin{figure*}
      \centering
\includegraphics[width=0.95\textwidth]{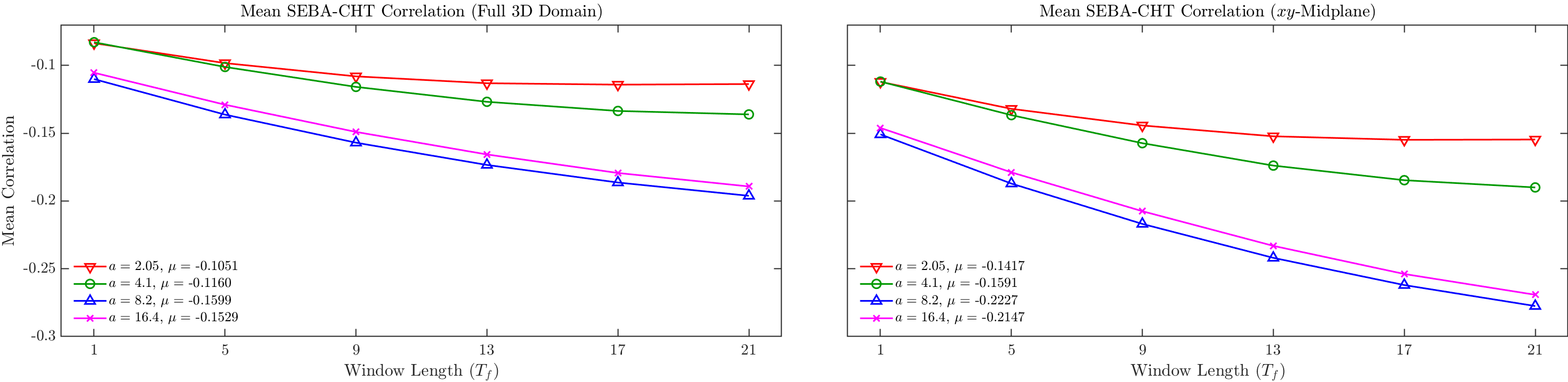}
      \caption{Plots of the mean correlation values (over the full time interval $[2001,2103] \ T_f$) between SEBA and CHT data over the entire three-dimensional spatial domain $M$ (left) and restricted to the $xy$-midplane (right) against the lengths of the time windows used to calculate the average CHT data. The colours (and symbols) in these plots represent the $a$ values used when executing the inflated generator method. Included in the plot's legend are the means of the mean correlation values $\mu$ taken over the window lengths for each curve.}
      \label{fig:rbc_cht_corr_mean}
\end{figure*}

This weak correlation observed between SEBA and CHT is likely due to the fact that we are computing these correlation values over the entire three-dimensional spatial domain $M$, and the correlation between these quantities is weaker within some parts of the domain compared to others, such as the thermal boundary layers along the top and bottom of the cell. To that end, we also compute the correlation between SEBA and CHT after restricting these data sets to the $xy$-midplane, with the graphs for these correlation values shown in the right-hand panel of Fig.\ \ref{fig:rbc_cht_corr}. 
Similar to the left panel in Fig.\ \ref{fig:rbc_cht_corr}, the time window over which we compute CHT averages changes, but the temporal diffusion parameter $a$ remains fixed at 4.1. We do not observe a great difference between the correlation values for the three-dimensional data and the $xy$-midplane data when CHT is defined on a single time step. However, as the length of the time window over which we compute the CHT averages increases, we see an improvement in these correlation values compared to those generated using the full three-dimensional data. These correlation values are still relatively weak as they roughly range between -0.25 and -0.1 (when CHT has been averaged). 

A likely reason for this lack of a stronger correlation is to do with the observation of a weaker qualitative connection between high SEBA and low-to-medium CHT values in the $xy$-midplane, as observed in Fig. \ref{fig:rbc_sebaAndcht_2052Tf}. For a stronger negative correlation between these two quantities (i.e. corresponding to a correlation value close to -1), quasi-stationary families of almost-invariant sets (identified through high values of SEBA) would have to align near perfectly with regions of $M$ corresponding to low to no convective heat transfer. At the same time, the correlation values are not zero because we do observe some qualitative connections between low SEBA and high CHT. The volume covered by these relatively thin filaments is much smaller compared to the volume covered by the metastable plumes of fluid identified through quasi-stationary families of almost-invariant sets and low CHT; hence this correlation between SEBA and CHT exists over a lower volume of $M$ and contributes to the correlation values being closer to 0 in Fig. \ref{fig:rbc_cht_corr}.

\subsection{Correlation between SEBA and other quantities}

To further test the possible connections between SEBA and the characteristics of our RBC flow system, we extend this analysis further by calculating the correlation between SEBA and several other key quantities for the RBC flow. To this end, we take correlations between SEBA and the following fields
\begin{itemize}
    \item Temperature fluctuation $\theta$, and its absolute value $|\theta|$;
    \item Vertical velocity $u_z$, and its absolute value $|u_z|$;
    \item Convective heat transfer (CHT) $H$, and its absolute value $|H|$;
    \item Magnitude of horizontal velocity $\| (u_x,u_y,0)^{\top} \|$;
    \item Magnitude of horizontal gradient of temperature fluctuation $\| (\partial \theta / \partial x,\partial \theta / \partial y,0)^\top\|$.
\end{itemize}
We average these quantities over subintervals of time of length 5 $T_f$ centred at $t_k$, $[t_k-2,t_k+2]$, as we did previously with the CHT data. Figure \ref{fig:rbc_cor_variousQuants} displays these correlation values across time for both the full three-dimensional spatial domain $M$ and the restriction to the $xy$-midplane. The mean correlation values across time are included in the legend of Figure \ref{fig:rbc_cor_variousQuants}. We observe particularly weak correlation between SEBA and the quantities $\theta$, $|\theta|$, $u_z$, $\| (u_x,u_y,0)^{\top} \|$ and $\| (\partial \theta / \partial x,\partial \theta / \partial y,0)^\top\|$ across the full three-dimensional domain. The correlation between SEBA and $|u_z|$ is somewhat less than between SEBA and $H$ and $|H|$;  therefore, we can be confident that our analysis focuses on CHT. 
%; especially in comparison to the correlation values calculated between SEBA and $H$ (CHT), indicating a lack of connection between our quasi-stationary families of almost-invariant sets (and the less metastable portions of the domain which separate these families of sets) and these quantities. 
%Restricting these quantities to the $xy$-midplane and re-calculating the correlations causes the correlation between SEBA and $\| (\partial \theta / \partial x,\partial \theta / \partial y,0)^\top\|$ to improve only slightly, causes the correlation between SEBA and both $\theta$ and $\| (u_x,u_y,0)^{\top} \|$ to get worse; and does not appear to have too much of a tangible effect on the correlation between SEBA and $u_z$.
%Taking the absolute values of $\theta$, $u_z$ and $H$; then computing the correlation between SEBA and these quantities over the full three-dimensional spatial domain $M$ only slightly improves these correlation values, with the most noticeable change observed with the vertical velocity $u_z$, with the mean correlation across time between SEBA and $u_z$ in the full spatial domain $M$ being -0.0020, and this considerably improving to -0.0535 when we take $u_z$; though this correlation value is still relatively poor in comparison (and on its own, as the correlation value is still close to 0). 
%Restricting these correlation computations to the $xy$-midplane, we see marginal improvements in the correlation values between SEBA and $|u_z|$ and SEBA and $|H|$. 
Restricting our calculations to the $xy$-midplane, greatly improves the correlation between SEBA and $|\theta|$, with the mean correlation changing from -0.0247 across the full spatial domain $M$ to -0.1720 across the $xy$-midplane. Restricted to the midplane, the correlation with $|\theta|$ exceeds the correlations with CHT. A reason for this observation might be that the strongest plume clusters, which have clear $|\theta|$ signatures, will be observed in the 
$xy$-midplane, while smaller-scale structures formed at the floor and ceiling may become washed out by turbulent mixing by the time they reach the $xy$-midplane. This was also observed in ref.\ \cite{Shevkar2025}.     

Finally, to benchmark against a featureless scalar field, we computed the correlations between SEBA and a constant scalar field, and between CHT and a constant scalar field.
The magnitudes of these correlations across all time windows shown in Fig.\ \ref{fig:rbc_cht_corr} were at least an order of magnitude smaller (not shown) than the values shown in Figs \ref{fig:rbc_cht_corr} and \ref{fig:rbc_cor_variousQuants}.

\subsection{Impact of the temporal diffusion parameter $a$}

To identify what sort of impact (if any) the temporal diffusion parameter $a$ used to construct the inflated generator has on the correlation results presented in the previous subsection, we re-run the inflated generator method for a smaller value of $a$ and two larger values of $a$ than the original value of 4.1 and perform similar correlation analysis using the SEBA vectors generated. As a convenient way of doing this, we re-run the inflated generator method for $a/2 = 2.05$, $2a = 8.2$ and $4a = 16.4$. The results are shown in Figure \ref{fig:rbc_cht_corr_mean}, where we calculate the mean correlation values between SEBA and CHT across the full time interval $[2001,2103] \ T_f$ and plot these against the lengths of the time windows over which we compute the CHT averages. Analogously with Fig. \ref{fig:rbc_cht_corr}, the left-hand plot shows correlation values computed for three-dimensional SEBA and CHT data, while the right-hand plot shows these values for SEBA and CHT data restricted to the $xy$-midplane. 

Between our two smaller values of $a$ (2.05 and 4.1), we see roughly no impact on the correlation between SEBA and single-time instances of CHT; though as the time window taken to compute the CHT averages increases in length, the mean correlation values improve for the larger $a = 4.1$ and plateau a little more quickly when $a = 2.05$. When $a = 8.2$, we find better mean correlation values when using single-time instances of CHT data, and the correlation values continue to improve as we average CHT over a larger time window. However, when we increase $a$ to 16.4, we do not observe any further improvements to our correlation results. Instead, while the mean correlation values for $a = 16.4$ are better than those for $a = 2.05$ and 4.1, these means are slightly worse (in the sense that they have value closer to 0, when we wish them to tend more towards -1) than those recorded for $a = 8.2$. Hence, increasing the value of the temporal diffusion parameter $a$ for the inflated generator does improve the correlation between SEBA and CHT but only up to a point, after which the correlation values either plateau or start to recede towards 0.

\begin{comment}
    From these two plots, we again ascertain that SEBA and CHT are correlated slightly more strongly on the $xy$-midplane compared to the entire three-dimensional spatial domain $M$, and increasing the length of the time window over which we average the CHT data also improves this correlation. However, as we can see from the two smallest $a$ values considered (2.05 and 4.1), the mean correlation in both Figures appears to plateau to a limiting value as the time window length increases, reaffirming that there is a limit as to how much we can improve the correlation values between SEBA and CHT by increasing the length of the time window over which we average the CHT data. We do not appear to have reached this limiting value yet for the two largest $a$ values (8.2 and 16.4), though the shapes of the blue and pink curves in both of these graphs suggest that we will eventually reach this limit for a larger time window length than those considered.
\end{comment}

\section{Conclusion}
%\gf{The conclusion is currently focussed around the correlation between the SEBA and CHT field, but there isn't any connection to the careful setup in the introduction that introduced large-scale transient patterns, turbulent superstructures, etc..., which was our main motivation. Can something be added to ``circle back'' and connect the conclusion to the main issues and questions outlined in the introduction?}
The main objective of our study was the application of a new method to detect and analyse quasi-stationary families of almost-invariant sets of convective heat transfer in a plane-layer Rayleigh-B\'{e}nard convection flow. In this flow, heat which is supplied at the bottom of the layer is carried to the top amplified by turbulent motion of the fluid layer between. The method that we used extract these families of sets is based on the so-called generator of the Perron-Frobenius or transfer operator that is inflated from space to space-time to capture the transient dynamics of the almost-invariant sets in time. 
The scalar field that was mainly analysed in our study is the convective heat transfer field $H({\bm x},t)$, which couples the vertical velocity component, $u_z$, and the temperature fluctuation, $\theta({\bm x},t)$. The quasi-stationary families of almost-invariant sets are obtained as eigenvectors of the discretized version of the inflated generator and a subsequent Sparse Eigenbasis Approximation (SEBA). 

In our three-dimensional data analysis, we  demonstrated qualitative and quantitative correlations between the patterns of convective heat transfer and the computed SEBA vectors and thus the capability of this analytical approach to extract the formation and dissipation of  transient almost-invariant sets in the complex dynamics of a turbulent convection flow. This implies that the mathematical framework, which we applied here, can describe the gradual transient dynamics of the large-scale patterns in the the plane-layer convection flow, the turbulent superstructures, and reveal its central role for the turbulent heat transfer. 

Possible extensions of this work for the future could be as follows. If the convection dynamics proceeds in a more complex settings, e.g., in a thin shell that is subject to rotation, then the transport patterns will become latitude-dependent and meridional transport might be suppressed. The transport barriers that would form in such a flow configuration should be identifiable by the present framework.

\section*{Acknowledgements}
The research of AB and GF is supported by the Australian Research Council through a  Discovery Project (DP2101003570) and a  Laureate Fellowship (FL230100088) respectively. 
The work of R.J.S. and J. S. is funded by the European Union (ERC, MesoComp, 101052786). Views and opinions expressed are however those of the authors only and do not necessarily reflect those of the European Union or the European Research Council.

\section*{Data Availability}
Code and data used in this work is available at \url{https://github.com/gfroyland/Inflated-Generator/tree/main/Rayleigh-Benard%20convection}.
%\gf{It is good that citations \cite{hcubature} and \cite{arnoldimethod} include the version number, but the appearance of the year at the end of the citation looks odd. Are these papers?  If so, the year goes with the paper, not the URL or code version number. If they are not papers, and simply code repositories, we don't need a year.}
%\bibliographystyle{plain}
\bibliography{citations}

\end{document}